\documentclass[manuscript]{emulateapj}
\usepackage{color}

\shorttitle{The impact of the dynamical state on the stellar population}
\shortauthors{M. Raouf et al.}

\definecolor{darkgreen}{rgb}{0.0,0.5,0.0}
\definecolor{darkred}{rgb}{0.5,0.0,0.0}
\definecolor{brown}{rgb}{0.65,.16,0.16}
\definecolor{grey}{rgb}{0.4,0.5,0.6}

\begin{document}
	
	\title{The impact of the dynamical state of galaxy groups on the stellar populations of central galaxies}
	
	\author{Mojtaba Raouf,\altaffilmark{1}
		Rory Smith,\altaffilmark{1}
		Habib G. Khosroshahi,\altaffilmark{2}
		Ali A.~Dariush,\altaffilmark{3}
		Simon Driver ,\altaffilmark{4}
		Jongwan Ko ,\altaffilmark{1}
		Ho Seong Hwang,\altaffilmark{1}
	}
	\affil{$^1$Korea Astronomy and Space Science Institute, 776 Daedeokdae-ro, Yuseong-gu, Daejeon 34055, Korea}
	\affil{$^2$School of Astronomy, Institute for Research in Fundamental Sciences (IPM), Tehran, 19395-5746, Iran}
	\affil{$^3$Institute of Astronomy, University of Cambridge, Madingley Road, Cambridge CB3 0HA, UK}
	\affil{$^4$International Centre for Radio Astronomy Research (ICRAR), The University of Western Australia, 35 Stirling Highway, Crawley, WA 6009, Australia}
	\email{raouf@kasi.re.kr}
	
	\begin{abstract}
		
		We study the stellar populations of the brightest group galaxies (BGGs) in groups with different dynamical states, using GAMA survey data. We use two independent, luminosity dependent indicators to probe the relaxedness of their groups; the magnitude gap between the two most luminous galaxies ($\Delta M_{12}$), and offset between BGG and the luminosity center ($D_{offset}$) of the group. Combined, these two indicators were previously found useful for identifying relaxed and unrelaxed groups. We find that the BGGs of unrelaxed groups have significantly bluer NUV-r colours than in relaxed groups. This is also true at the fixed sersic index. We find the bluer colours cannot be explained away by differing dust fraction, suggesting there are real differences in their stellar populations. SFRs derived from SED-fitting tend to be higher in unrelaxed systems. This is in part because of a greater fraction of BGGs with non-elliptical morphology, but also because unrelaxed systems have larger numbers of mergers, some of which may bring fuel for star formation. The SED-fitted stellar metallicities of BGGs in unrelaxed systems also tend to be higher by around 0.05 dex, perhaps because their building blocks were more massive. We find that the $\Delta M_{12}$ parameter is the most important parameter behind the observed differences in the relaxed/unrelaxed groups, in contrast with the previous study of Trevisan et al. (2017). We also find that groups selected to be unrelaxed using our criteria tend to have higher velocity offsets between the BGG and their group.

	\end{abstract}
	
	\section{Introduction}
	
	Investigating the evolution of galaxies in different environments over cosmic time is one of the requirements to improve our understanding of the galaxy formation process. In the local universe, there is a strong anti-correlation in galaxy populations, between galaxy star formation rate and their environmental density, with passively evolving early-type galaxies preferentially found within the dense cores of massive galaxy clusters. In contrast, star-forming galaxies are preferentially found in lower density regions such as groups or the field \citep[e.g.,][]{Dressler1980}.
	
	Central galaxies in galaxy clusters are typically quiescent early-types with no significant star formation, as described by the morphology-density \citep{Dressler1980} and star formation-- density \citep{Dressler1985, Poggianti1999} relations. However, some ultraviolet (UV) and mid-infrared (IR) observations report low-level recent star formation in some optical-red-sequence objects \citep{Yi2005,Rawle2008, Ko2013}. Furthermore, far-IR observations of galaxies in small groups indicate that galaxies accreted onto clusters along with filamentary structures, where gravitational interactions between the galaxies, rather than cluster-potential, stimulate starbursts \citep{Fadda2008, Koyama2008,Haines2011}. It has been shown that recent star formation is well traced by both near-UV ($\lesssim$ 1 Gyr) and mid-IR ($\lesssim$ 2 Gyr) in quiescent, red, early-type galaxies \citep{Ko2013}. 
	
	It is useful to compare the stellar populations of central galaxies in relaxed and unrelaxed groups because of their differing time scales since their last major merger.  For instance, in relaxed galaxy groups in the most case, the recent major merger occurs earlier compared to unrelaxed systems \citep{Jones2003,Raouf2018}.     
	In order to understand \textit{how} the stellar populations of central galaxies are linked to their group halo's dynamical state, we focus on dynamically relaxed and unrelaxed groups. We define a parameter space which consists of optically measurable parameters of group galaxies, that allows an efficient age-dating of galaxy group growth history, characterizing them into relaxed (old) and unrelaxed (young) subcategories \citep{Raouf2014}.
	The first of the two independent optically measurable indicators to probe the dynamical state of the group halo is the magnitude gap between two most bright galaxies in group. This is expected to develop as a result of the internal mergers within groups, as argued in the formation of fossil galaxy groups\footnote{Fossil groups are characterized by the stellar dominance of the Brightest Group Galaxy (BGG) and thus have a large magnitude gap ($>$2 mag)between the two most luminous galaxies within a radius of 0.5$R_{\rm vir}$ and  $L_{X,bol}\approx 10^{42} $~h$_{50}^{-2}$ ~erg~s$^{-1}$ \citep{Jones2003}} \citep{Ponman1994} and demonstrated in cosmological simulations \citep{Dariush2010}. The second of the two independent optically measurable indicators to probe the dynamical state of the group halo is the offset between BGG location and the group center. Physcially this makes sense because BGGs are located at the peak of the X-ray emission in a relaxed system. Therefore, deviation in the location of the BGG from the group center can occur in merging systems or dynamically unrelaxed groups \citep{Smith2005,Smith2010,Rasmussen2012}.
	
	Observationally, \citet{Khosroshahi2017} found that the radio luminosity of the BGGs strongly depends on their dynamical age, such that the BGGs in dynamically unrelaxed groups are an order-of-magnitude more luminous in the radio than those with a similar stellar mass but residing in dynamically relaxed groups.  The presence of a large luminosity gap points at the absence of a recent major merger which could ignite cold mode accretion. This finding is consistent with results from hydrodynamical simulations \citep{Raouf2016}, which suggest that the black hole accretion in the BGGs of dynamically relaxed groups is lower for a given stellar mass than in unrelaxed groups.
	In \cite{Raouf2018}, the impact of the last major merger of central galaxies, within dynamically old and young groups of galaxies, has been shown on their associated 1.4 GHz AGN radio emission, as predicted by the {\sc Radio-SAGE} semi-analytic model of galaxy formation \citep{Raouf2017,Raouf2019}. This study suggests that the radio luminosities of central galaxies are enhanced in halos that assembled more recently like unrelaxed groups, independent of the time since their last major merger. 
	
	\citet{Trevisan2017} found no correlation between magnitude gap and BGG ages, metallicities, [$\alpha$/Fe], and star formation history using a Sloan Digital Sky Survey \citep[SDSS, ][]{York2000} -based sample of 550 groups with elliptical BGGs.  They suggest that BGGs in fossil groups have undergone more mergers than those in non-fossil groups, but these mergers are either dry or occurred at a very high redshift, which in either case would leave no detectable imprint in their spectra. They also show that Second Brightest Group Galaxies (SBGGs) in fossils lie significantly closer to the BGGs (in projection) than galaxies with similar stellar masses in normal groups, which appears to be a sign of the earlier entry of the former into their groups. Nevertheless, the stellar population properties of the SBGGs in fossils are compatible with those of the general population of galaxies of similar stellar masses residing in normal groups.
	
	The removal of the gas supply necessary for continued star formation could take place through a variety of mechanisms. Suggested gas sweeping processes have included galaxy-galaxy collisions \citep{Spitzer1951}, tidal encounters \citep{Toomre1972, Richstone1976}, ram pressure stripping \citep{Gunn1972,Park2009}, and removal of the external gas reservoir which is thought to be a crucial source of fuel for future extended star formation in late-type spiral galaxies \citep{Larson1978, Larson1980, Gunn1982}.
	Assuming a simple picture for the gravitational collapse of gas into the center of a galaxy group, BGGs located at the group center will be influenced by a larger reservoir (density) of hot gas compared to BGGs with a large offset. It seems the relaxed groups in which the BGGs have a smaller deviation from the group center contain more hot gas with respect to the unrelaxed groups. Such relaxed groups have no experience of the dry major merger in their formation history \citep{ Khosroshahi2004,Khosroshahi2006a,Khosroshahi2007}.
	The focus of our study is the relation of the BGG stellar population properties to their group halo's dynamical state, based on optically measurable parameters including the magnitude gap and BGG offset. We try to address the questions of how do the BGGs in groups of different dynamical states differ, in terms of their morphology, stellar metallicity, and star formation activity.
	The structure of the paper is as follows: the data and sample selections are described in Section 2. In Section 3, we present our results and analysis. Finally, we provide a summary and discussion in Section 4. Throughout this paper, we adopt $H_0 = 100h\ km\ s^{-1}Mpc^{-1}$ for the Hubble constant with $h = 0.7$.
	
	\begin{figure}
		\includegraphics[width=0.49\textwidth]{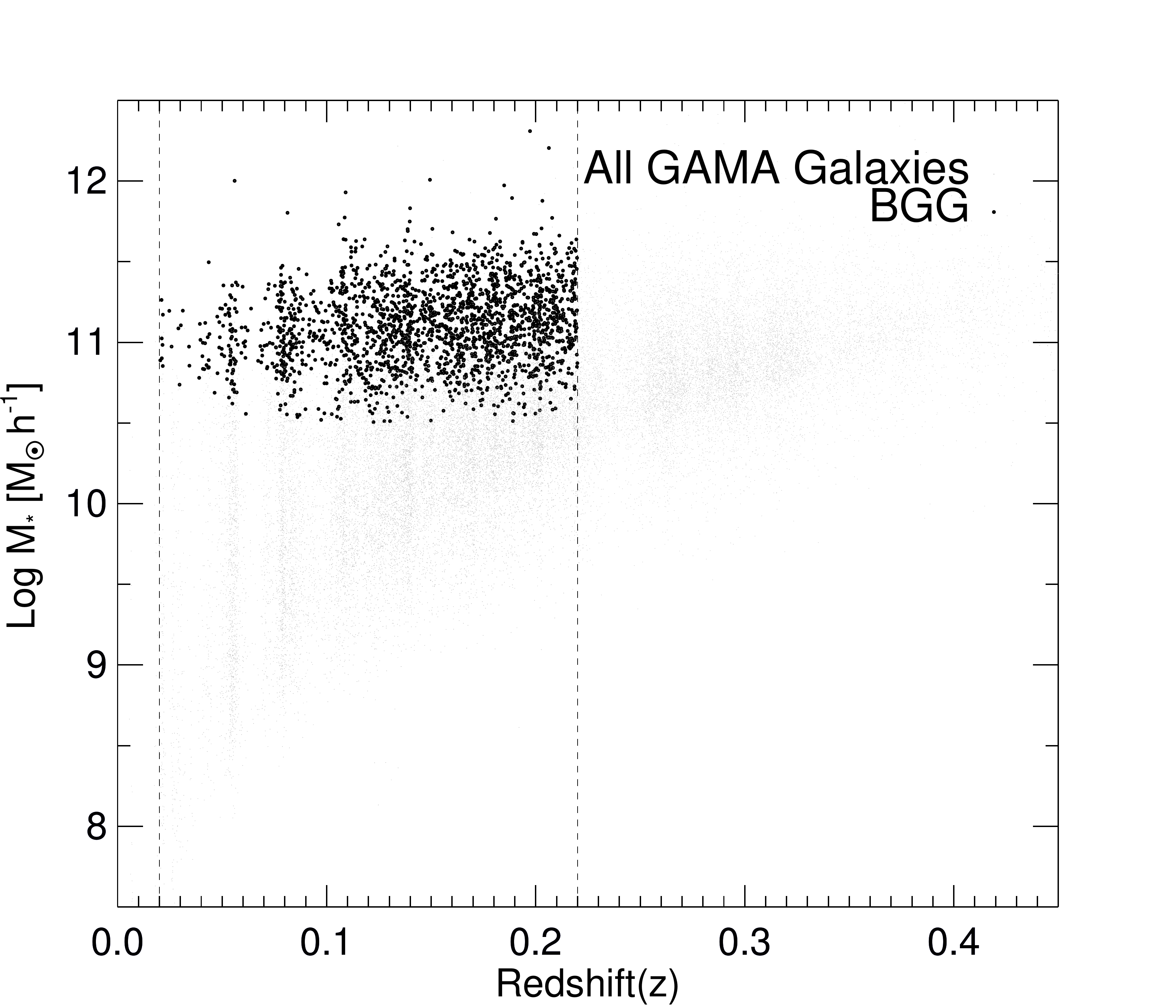}
		\caption{Sample selection; the stellar mass of the BGGs as a function of the redshift. The background grey dots represent all galaxies assigned to groups in the entire GAMA database. The black dots represent our sample of luminous BGGs ($M_r\leq-21.5$) within the redshift limit which is defined based on the redshift completeness of our sample.}
		\label{fig:redshift_distribution}
	\end{figure}
	
	\begin{figure*}
		\centering
		\includegraphics[width=0.33\linewidth]{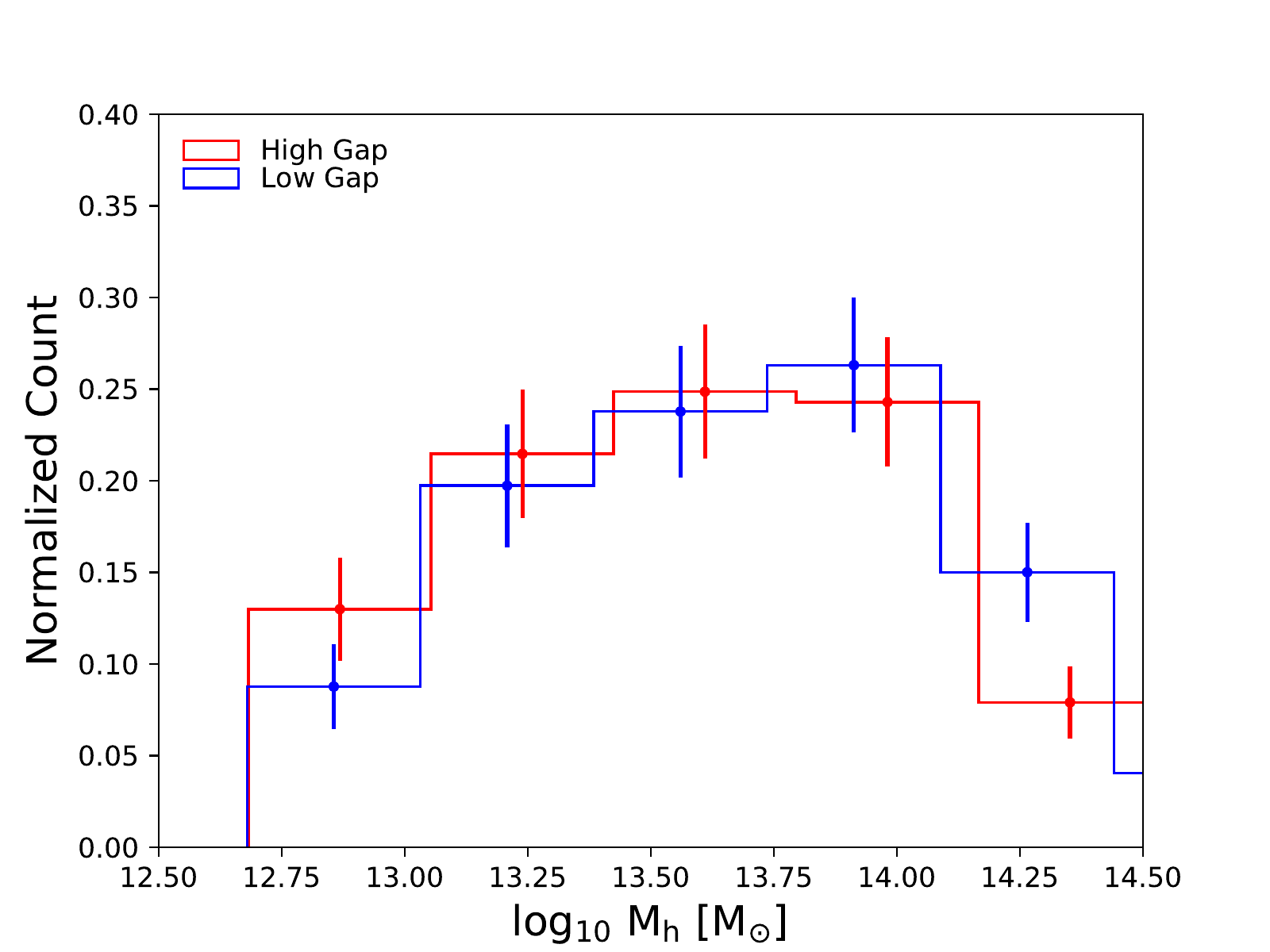}
		\includegraphics[width=0.33\linewidth]{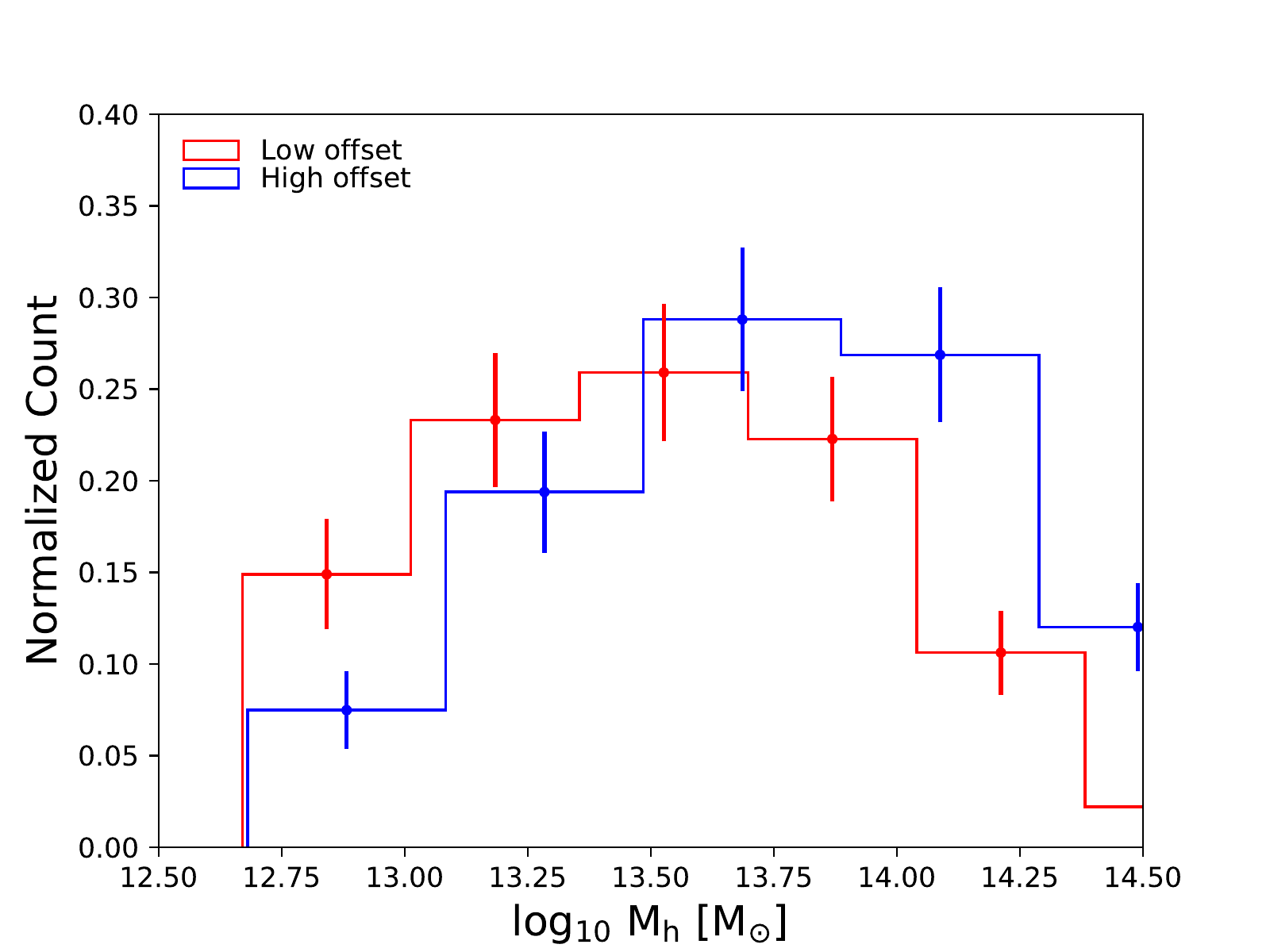}
		\includegraphics[width=0.33\linewidth]{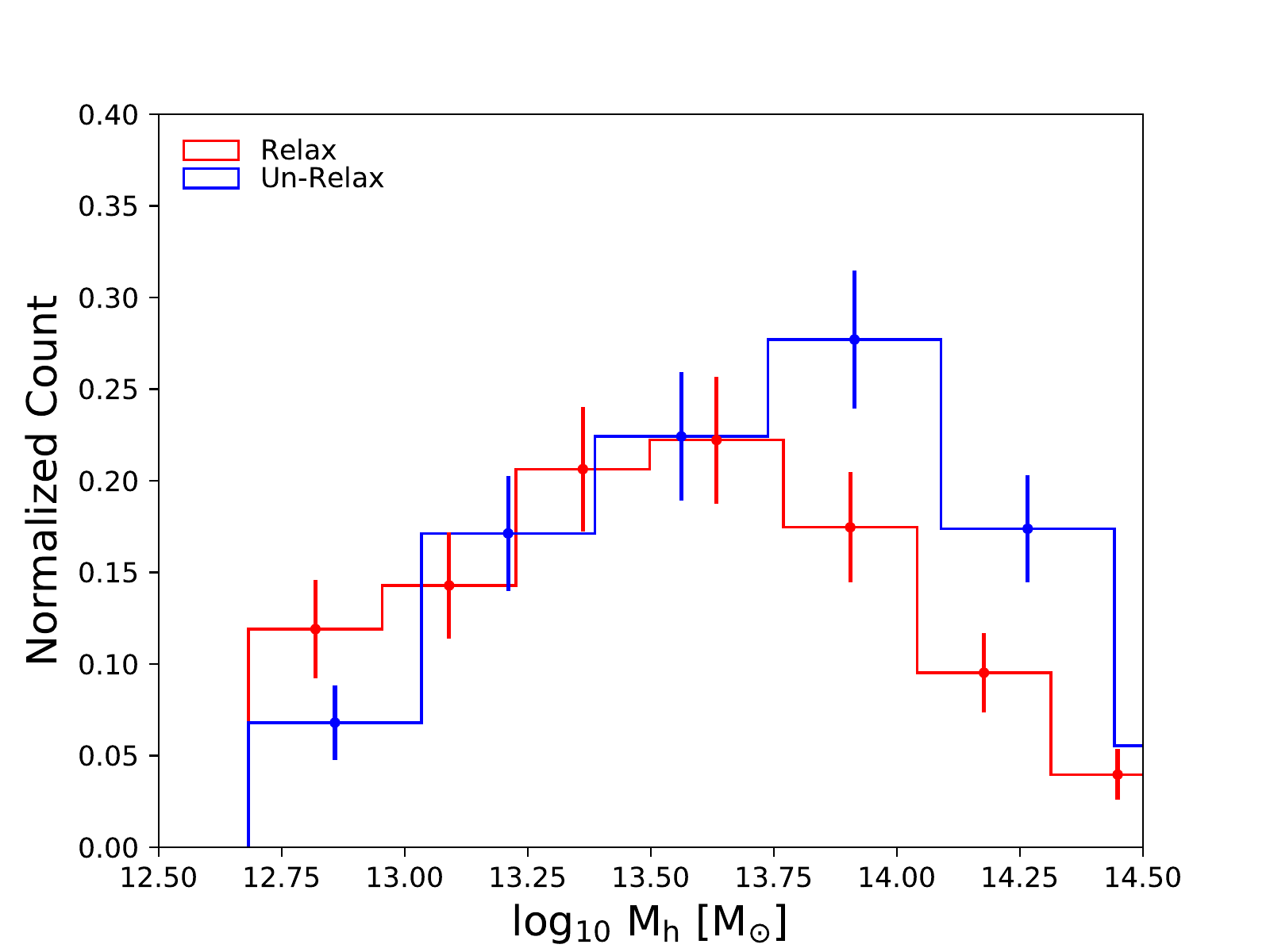}
		\includegraphics[width=0.33\linewidth]{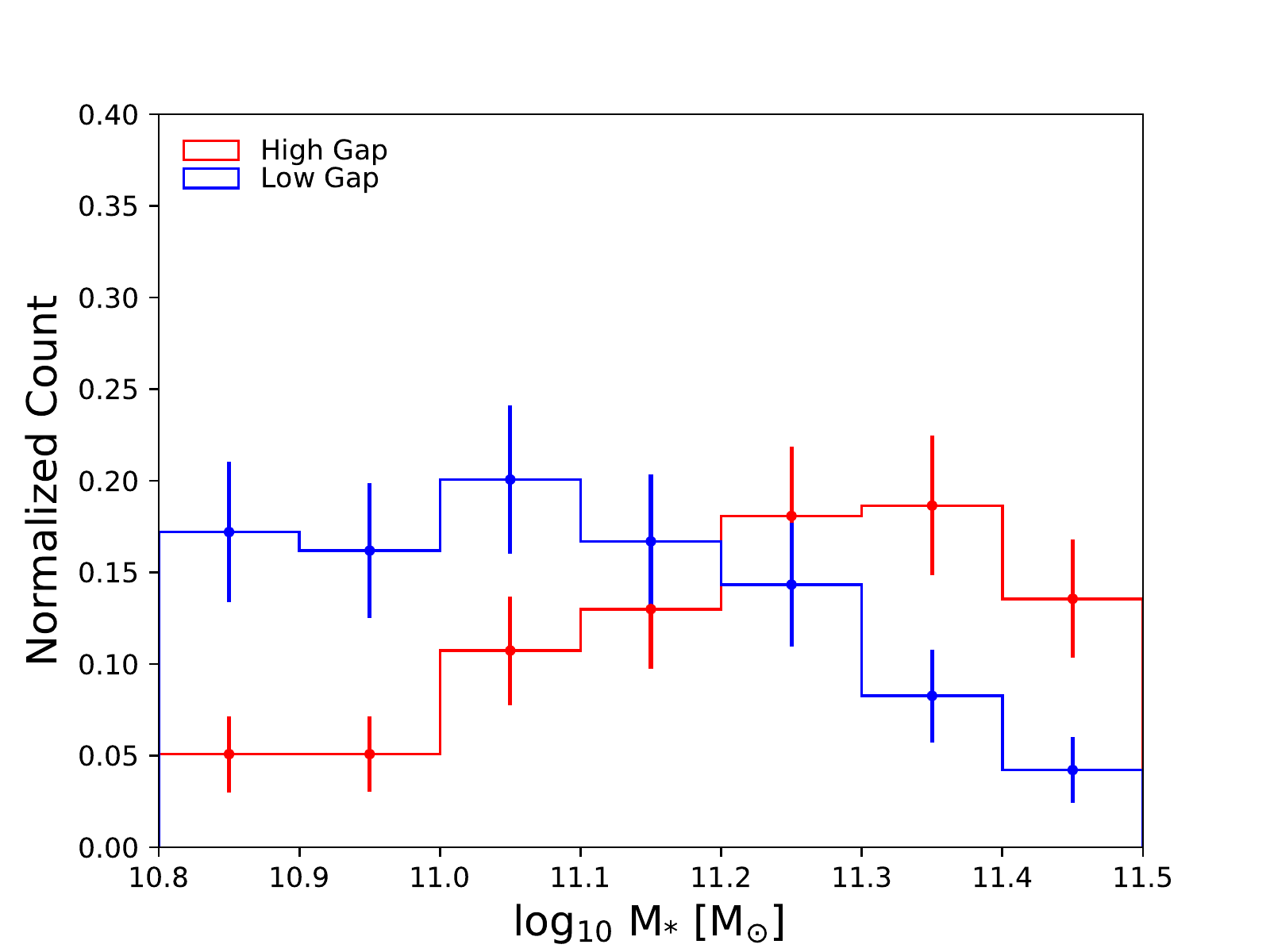}
		\includegraphics[width=0.33\linewidth]{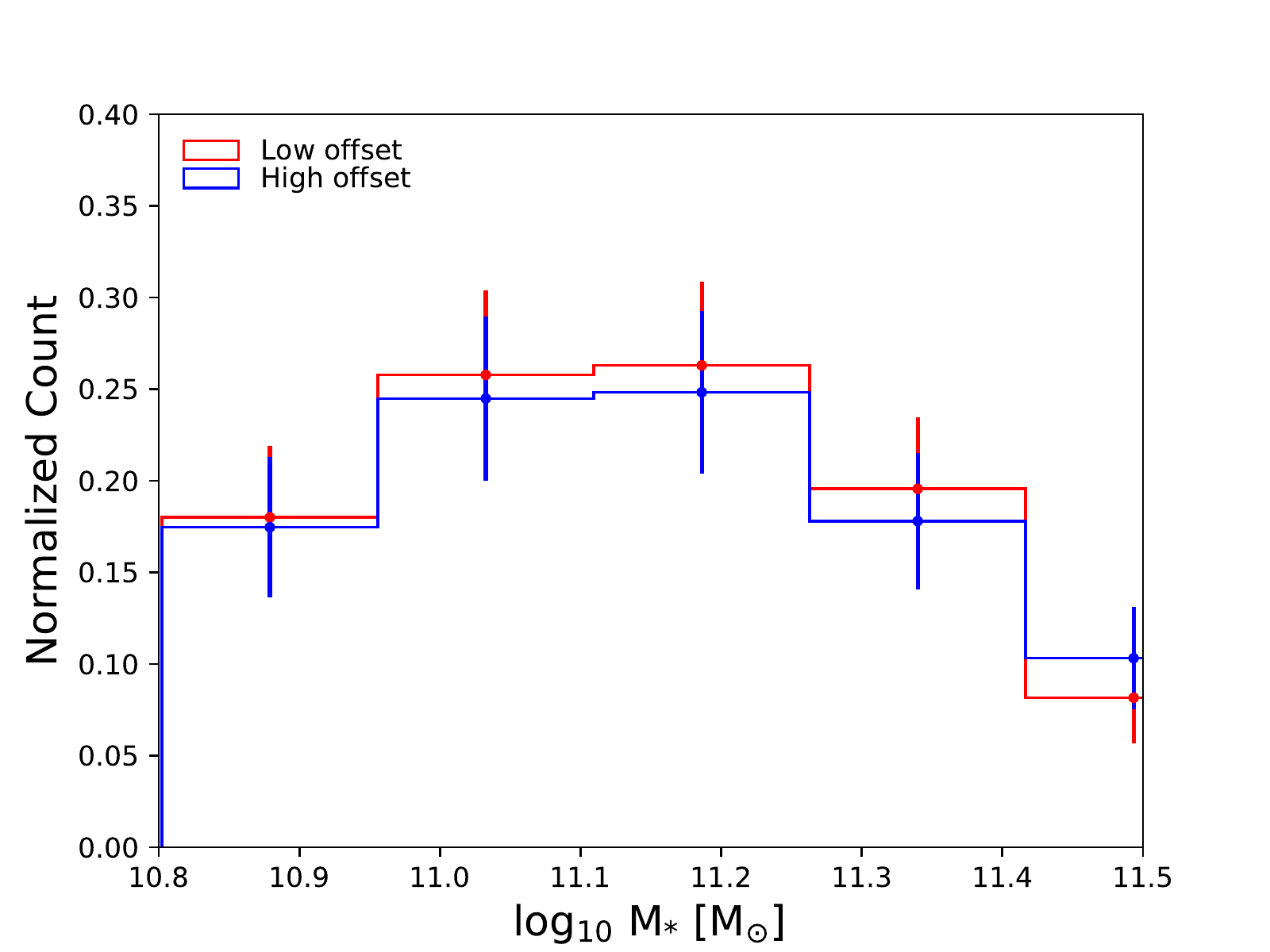}
		\includegraphics[width=0.33\linewidth]{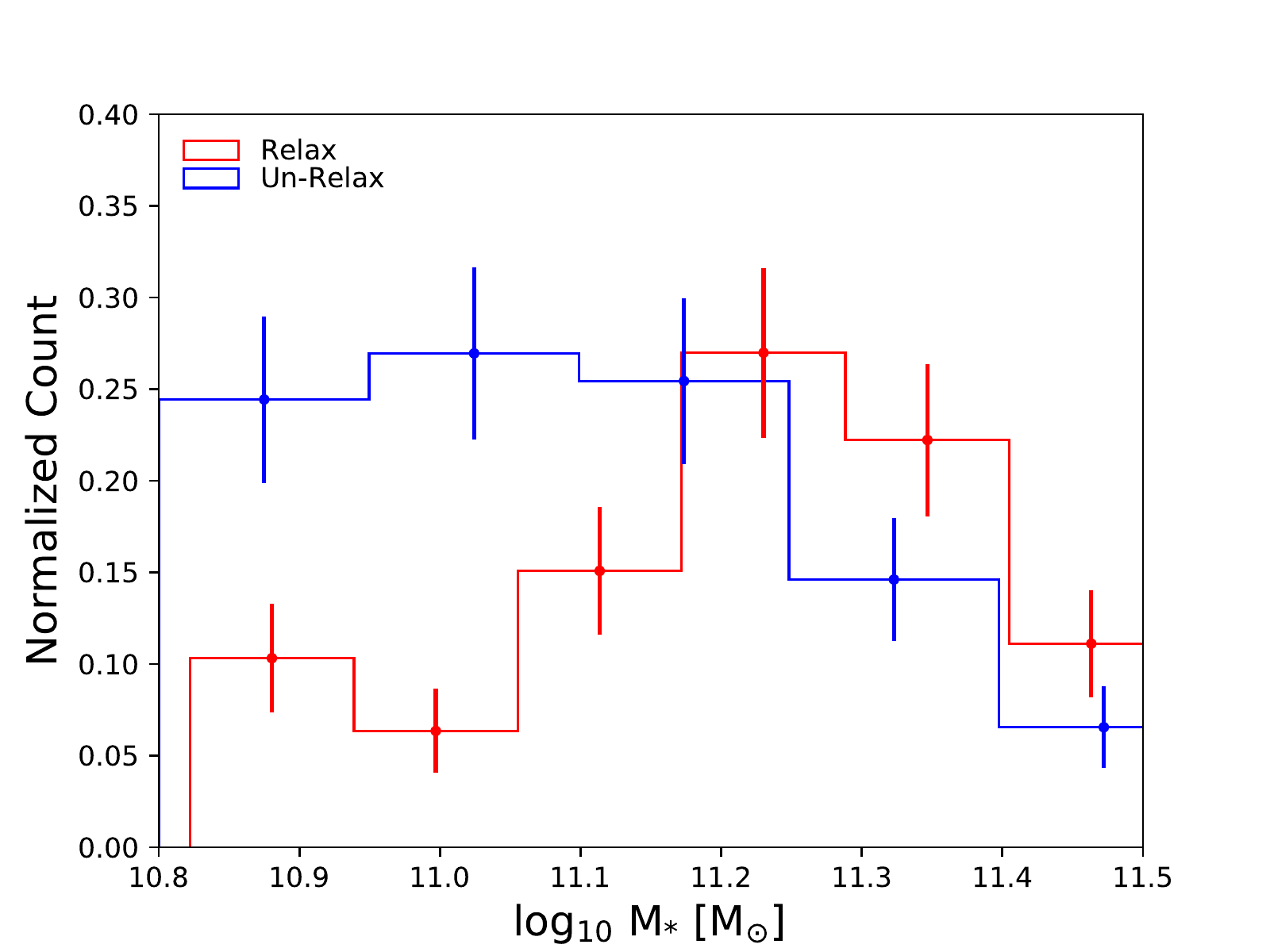}
		\caption{Distribution of halo mass(top) and BGG stellar mass (bottom) with Poisson error bars for comparing all sub-samples including relaxed/unrelaxed, high/low gap and low/high offset as shown by red/blue color in each panel. }
		\label{fig:nuv-hist-mh-ms}
	\end{figure*}
	
	\section{Data and sample selection}
	The main source of data for this study is the Galaxy And Mass Assembly (GAMA) survey, a multi-wavelength spectroscopic data set covering an area of 180 deg$^2$ in three distinct equatorial regions. The description of the survey input catalogue selection is given in \cite{Baldry2010}, while other aspects of the survey have been described in \cite{Robotham2010}, \cite{Driver2011} and  \cite{Hopkins2013}. 
	
	We use the third data release, GAMA-DRIII (see \cite{Baldry2010} or http://www.gama-survey.org/dr3 ). In brief, GAMA target selection is based on data from the Sloan Digital Sky Survey Data Release 6, and the input catalogue was selected to include all galaxies with Galactic extinction corrected Petrosian magnitudes above $19.8$ mag. Additional $(J-K)$ near-IR colour selection was used to include possible AGN as well as size selection to recoup extended objects misclassified as stars. In \cite{Liske2015} it was demonstrated that the spectroscopic sample is over 98 per cent complete, and unbiased in colour, size, clustering, magnitude or surface brightness within the selection limits (i.e., $15.0 < \mu_{r,50} < 26.0$ mag/sq arcsec).
		In addition to the photometry and flux measurements, we also make use of the GAMA Galaxy Group catalogue, as described in \cite{Robotham2011}. The group catalogue is constructed using a friends-of-friends algorithm calibrated against mocks constructed from the Millennium Simulation \citep{Springel2005}.

	The catalog contains 23,838 galaxy groups which reduce to about 4,000 galaxy groups and about 19,000 group members with a multiplicity of at least 4 spectroscopically confirmed members and within the limit of our choice of redshift range (Figure \ref{fig:redshift_distribution}). We also consider only the brightest group galaxies brighter than $M_r=-21.5$ mag. This choice maximizes the size of our sample within our redshift range in terms of having a complete set of groups in which the first and second most luminous galaxy are detectable above the GAMA luminosity limit of $r$=19.8. For example, if we remove the luminosity cut, we must reduce the maximum redshift ($z$ = 0.22) to $z$=0.09 to ensure completeness, but then the sample is reduced to less than 500 groups of galaxies. In the process, we tend to exclude modest Milky-way like galaxies as hosts of groups. This limit reduces our sample to 1654 galaxy groups.  Using the total extrapolated luminosity, and the total stellar mass of the group galaxies and their positions, we obtain the luminosity gap and the luminosity centroid of the groups.
	
	Our galaxy groups are also split into a dynamically relaxed and unrelaxed subsample using the following criteria:
	
	Criteria I. Galaxy groups with a large luminosity gap between the BGG and the second brightest group member, $\Delta M_{12} \ge 1.7$ (``high gap'') in r-band. In addition, we also impose that the BGG is located within a radius of 70 kpc of the luminosity/stellar-mass centroid of the group (``low offset''). This criteria reduces our subsample to 139 galaxy groups, labeled as ``relaxed" systems.
	
	Criteria II. Galaxy groups with a small luminosity gap, $\Delta M_{12} \le 0.5$ (``low gap'') in r-band. We impose the BGG to be located outside the radius of 70 kpc centred on the luminosity/stellar-mass centroid of the group (``high offset''). This reduces our subsample to 399 galaxy groups labeled as ``unrelaxed" systems. 
	
	Note that the small difference between the adopted $\Delta m_{12}=1.7$ limit used for the selection of the relaxed and high gap groups and the one conventionally used in previous studies of optical fossil groups, $\Delta m_{12}=2.0$, is to ensure a statistically meaningful number of galaxies in both the above samples. Other authors have also adapted similar variations in the sample selection of fossil galaxy groups \citep[e.g.,][]{Gozali14}. We also follow the conventional definition of low gap groups ($\Delta M_{12} \le 0.5$) based on the results of cosmological simulations \citep[e.g][]{Dariush2010}. 
		Our choice of an offset criteria of 70~Kpc was found to successfully divide relaxed and unrelaxed groups in cosmological simulations \citep[see fig.3 in][]{Raouf2014}.
	\begin{figure*}
		\centering
		\includegraphics[width=1\linewidth]{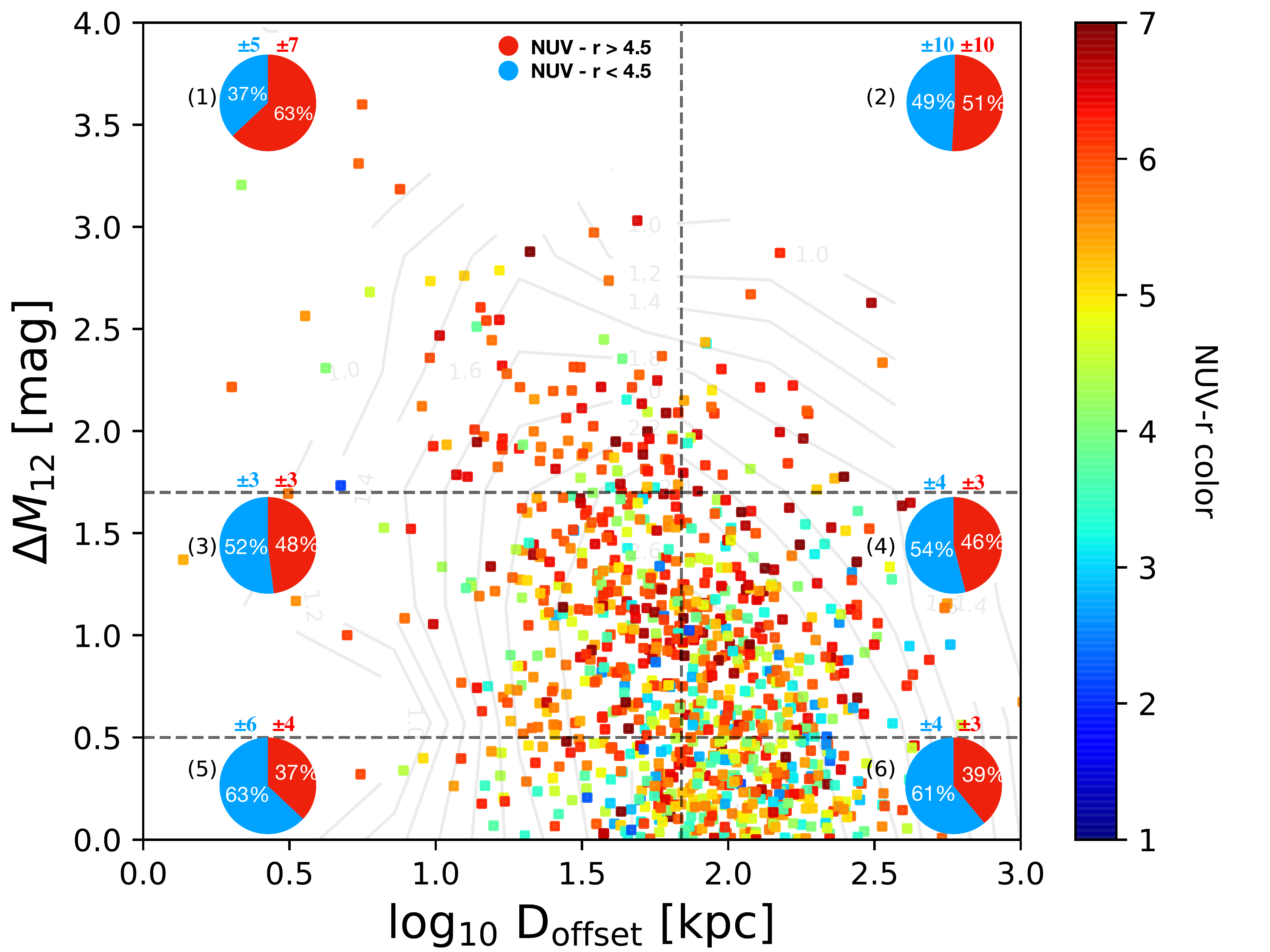}
		\includegraphics[width=0.49\linewidth]{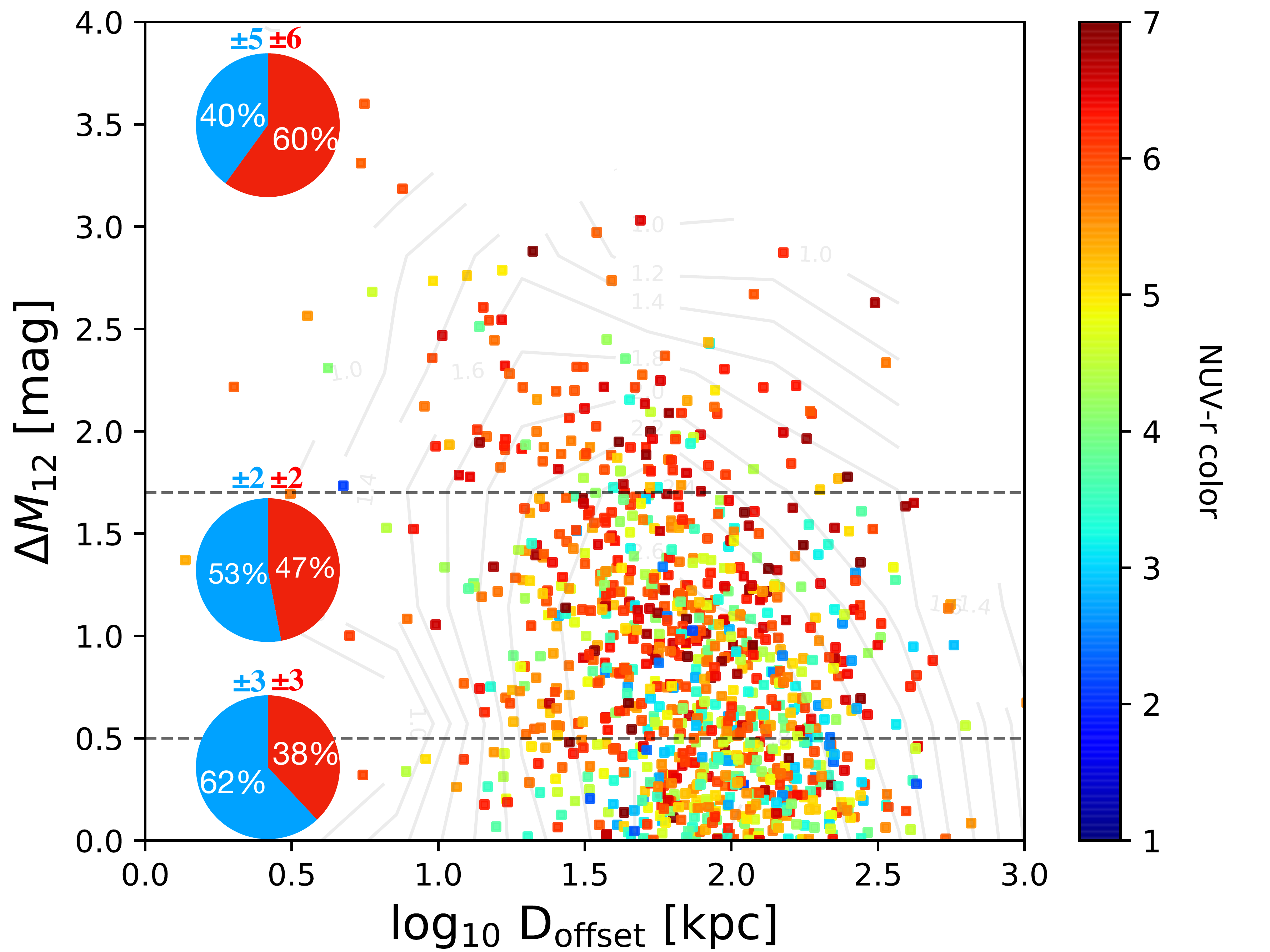}
		\includegraphics[width=0.49\linewidth]{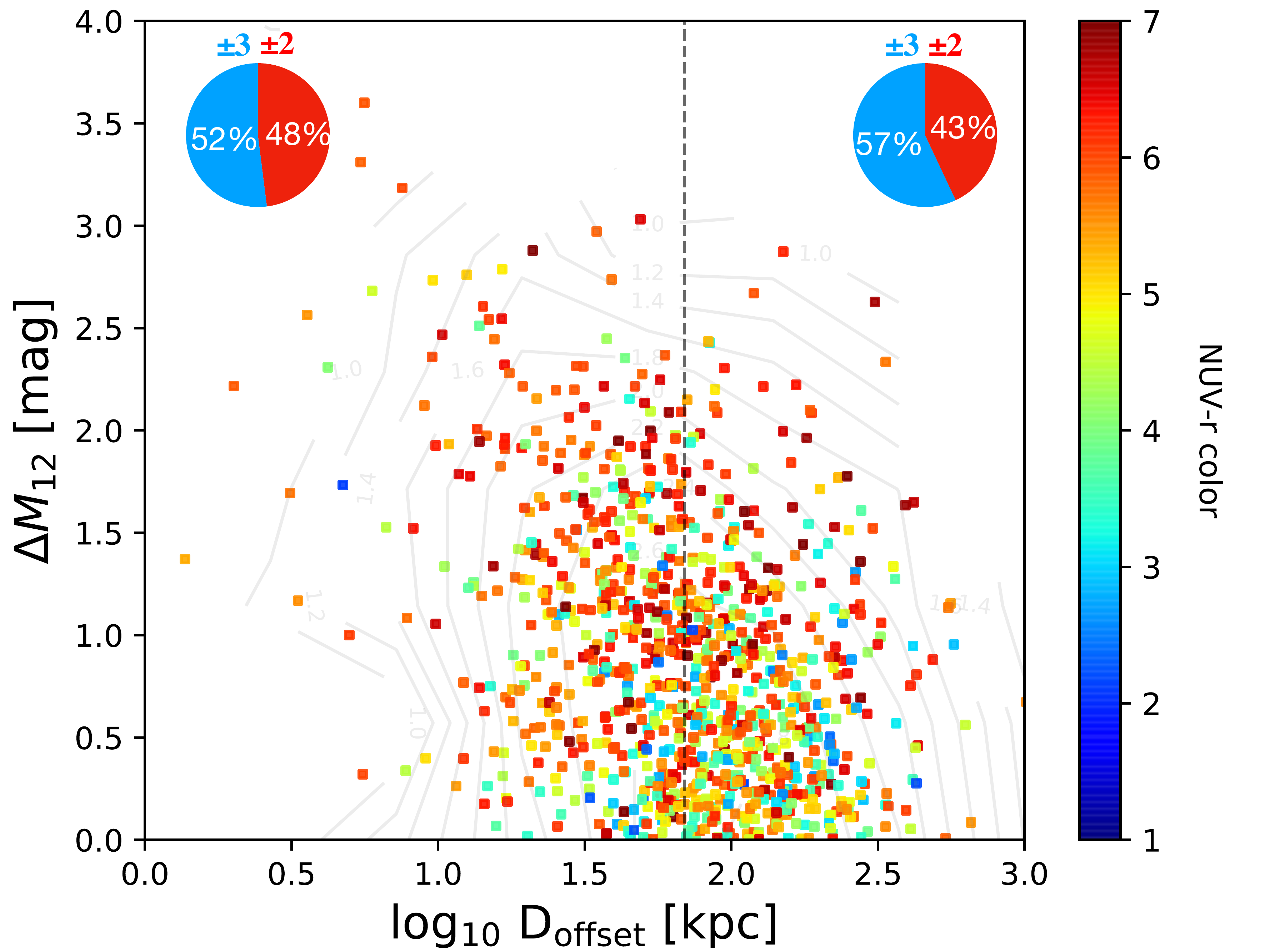}
		\caption{Distribution of  $\Delta M_{12}$ as function of $D_{offset}$ color code by the NUV - r color. Black dashed lines separate the regions for relaxed and unrelaxed systems region (1) ($\Delta M_{12} > 1.7$ and  $log(D_{offset}) < 1.8$) and region (6) ($\Delta M_{12} < 0.5$ and  $log(D_{offset}) > 1.8$), respectively. Red and blue regions of the pie-charts represent BGGs with NUV-r colours $>$4.5 or $<$ 4.5, respectively including the statistical Poisson errors for each regions.
			The \emph{grey contours} show the number density of galaxies per pixel (after smoothing by a Gaussian with $\sigma$ = 1.0 pixel). In the bottom panels, we show the same figure but for regions of low, intermediate and high gap (bottom left) and high and low offset (bottom right).}
		\label{fig:lum_gap_NUV}
	\end{figure*}
	\begin{figure}
		\includegraphics[width=1.05\linewidth]{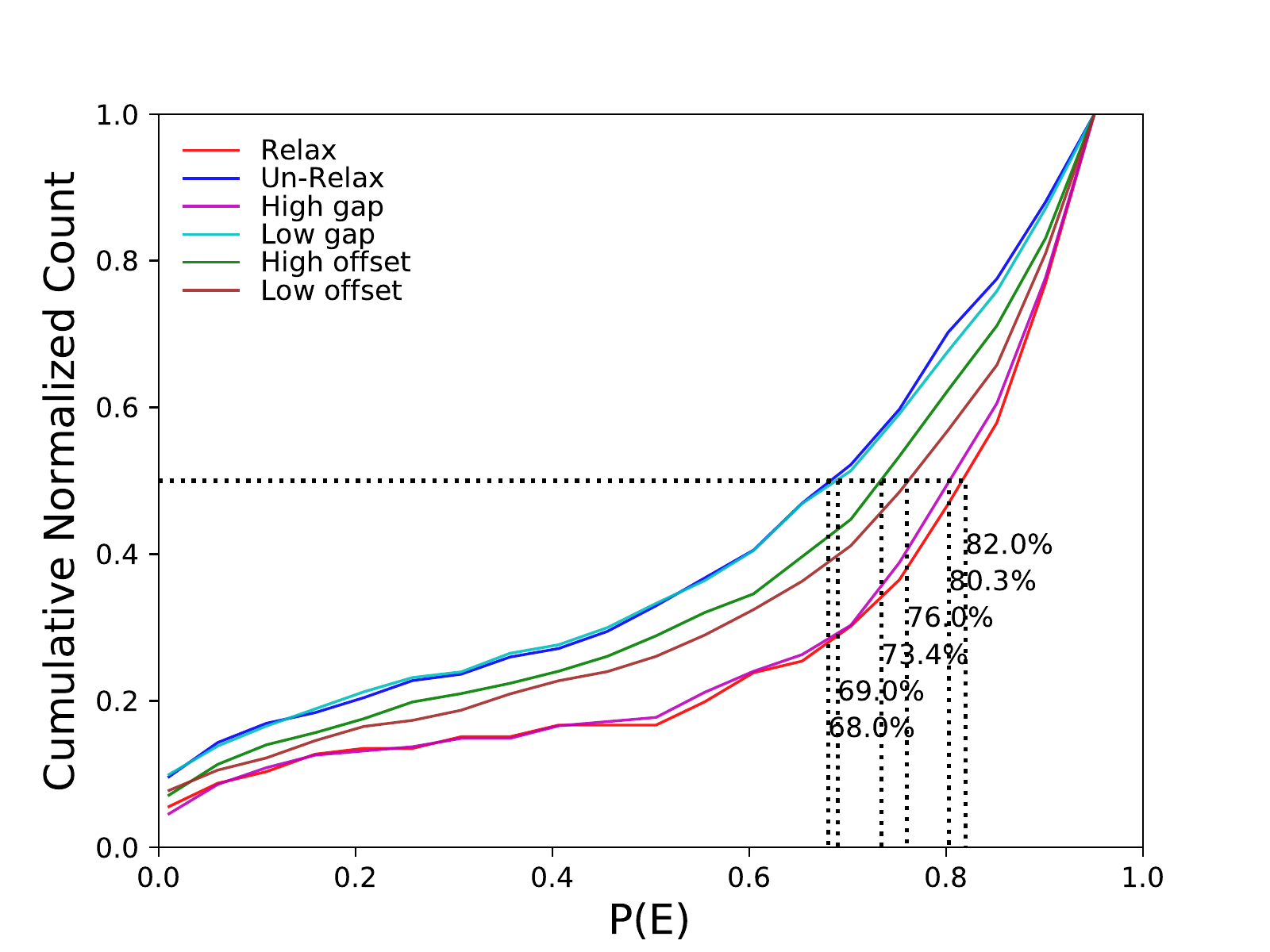}
		\caption{ Cumulative distribution of elliptical probability, P(E), for BGGs in different sub-samples of relaxed, unrelaxed, high/low gap and low/high offset galaxy groups. The dashed lines and percentage show the median elliptical probability for that sub-sample.}
		\label{fig:pellipgama}
	\end{figure}
	
	\begin{figure}
		\includegraphics[width=0.53\textwidth]{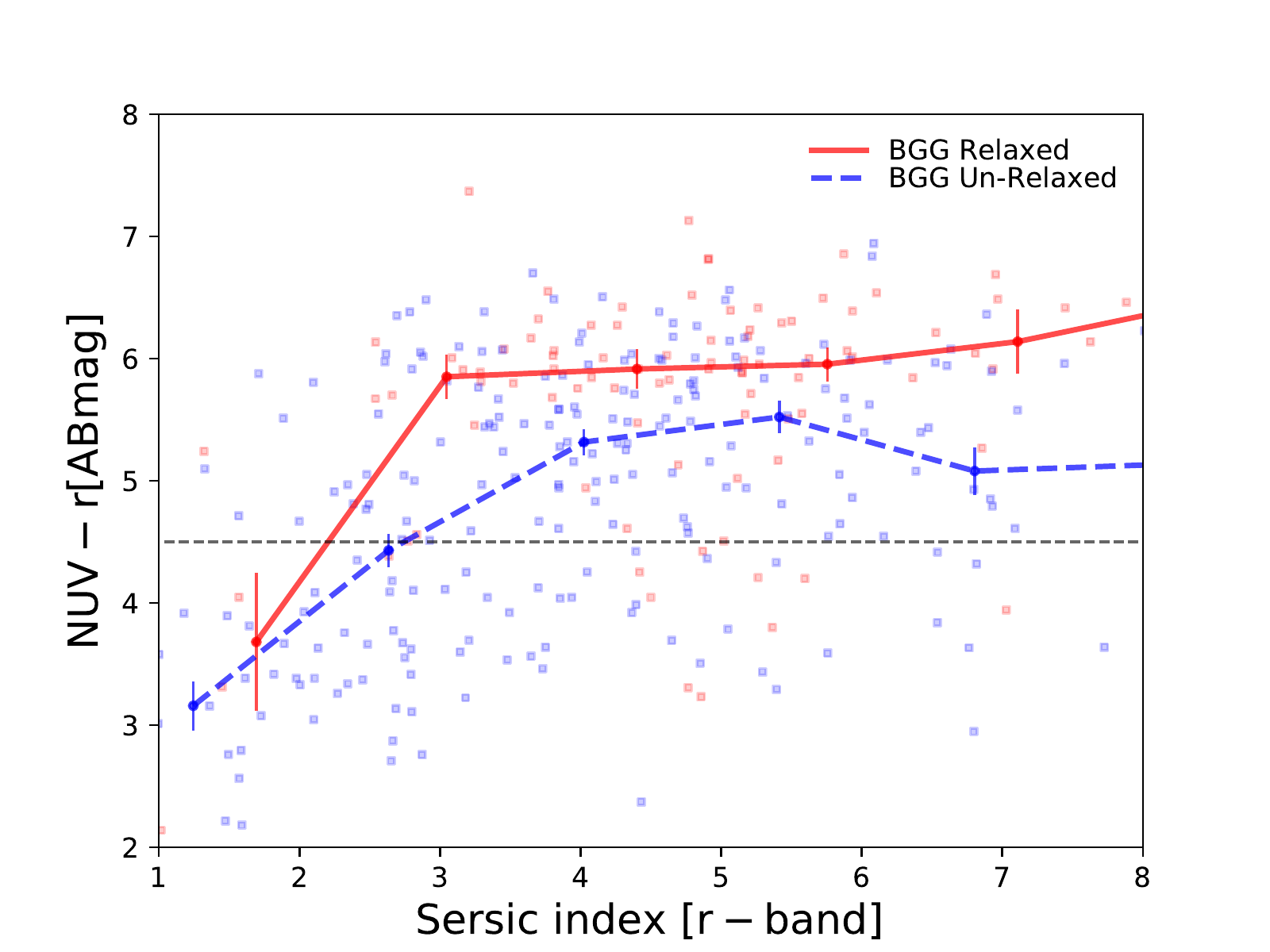}
		\caption{Distribution of NUV-r as a function of BGG's sersic index in r-band for relaxed (red) and unrelaxed (blue) samples. The \emph{red line} and \emph{blue dashed-line} show the medians and $\sigma/\sqrt{N}$  uncertainties for BGGs of relaxed and unrelaxed groups, respectively. The horizontal dotted-line shows the NUV-r=4.5 boundary used to divide the sample into red and blue galaxies.
		}
		\label{fig:NUV_flux_fossil_ctrl_scater}
	\end{figure}
	
	The luminosity centroid of the group members is provided in the GAMA group catalogue and is defined as the center of light derived from the r-band luminosity of all the galaxies identified to be within the group \citep{Robotham2011}. Our choice of redshift range (0.02 - 0.22) is chosen based on providing a complete sample of groups with a luminosity gap of $1.7$ mag (Figure \ref{fig:redshift_distribution}).
	
	In this work we use the stellar masses, SFRs, stellar metallicity and dust mass estimates from \citet{Driver2018}. These were derived using the energy balance SED fitting code magphys \citep{daCunha2008}, which fits the observed FUV-far-IR SEDs \citep[see][]{Wright2016} with UV/optical/NIR spectral templates of stellar populations and MIR/FIR templates of dust emission generated from \cite{Bruzual2003} and \cite{Charlot2000}, respectively, and assuming a Chabrier IMF \citep{Chabrier2003}. The code then determines the overall best fit stellar+dust template pair (regressing against the photometry) and outputs the physical parameters associated with these best-fit stellar properties. In this study, we use the values of the best-fit templates.
	
	We also note that the stellar metallicity estimates are not expected to be very accurate as they are based on SED-fitting alone. Furthermore, SFRs derived by magphys can not reach a value of zero due to the parameterized shape of the SFH. Nevertheless, they are still useful as we look for differences between our subsamples.
	In this study, we use NUV fluxes from the Galaxy Evolution Explorer {\sc{GALEX}} survey for estimating each galaxy's NUV-r color, which has a maximum error of 0.3 dex for our sample. We broadly split our sample of BGGs into two subsamples based on the NUV-r colour. Galaxies with NUV-r $>$ 4.5 and NUV-r $<$ 4.5 are described as a commonly used distinction between passive(red) and star-forming(blue) galaxies \citep[e.g.][]{Salim2007, Haines2011a,Rasmussen2012}.
	
	We also use the sets of filters provided by the mid-IR Wide-field Infrared Survey Explorer {\sc{WISE}} survey, which includes w1, w2, w3 and w4 corresponding to 3.4, 4.5, 12 and 22-micron data. Meanwhile, about 97\% of our sample has a w4 (22 Micron) detection from the {\sc{WISE}} survey. We note that more than half($\sim$55\%) of our total sample has FIR measurements in the form of the {\sc{Herschel}} survey catalogue.  
	
	The normalized distribution of halo mass and stellar mass for each subsample are shown in Figure \ref{fig:nuv-hist-mh-ms}(See Table \ref{tab:value} for the group number counts).
	We use dynamical group mass estimates based on the group velocity dispersion. We note a lower multiplicity can increase the erros on group mass and group centroid \citep[see sec.4.3][]{Robotham2011}. However we restrict our sample to have 4 or more members and, in fact, less than 20\% of our sample have a multiplicity of 4. We also restrict our sample to groups with masses above $10^{12.7} M_{\odot}$.

	The figure shows that our various subsamples (high/low gap, low/high offset and relaxed/unrelaxed groups) have a fairly similar range of halo mass and stellar mass. Thus most of the differences we find between our subsamples are not driven by differences in the halo and stellar mass. 
	For instance, increasing the halo mass limit to $10^{13} M_{\odot}$ and/or the group multiplicity to $>$ 5 on average leads to a decrease in the statistics of around 30\% in each sub-sample, arising from the scatter, but with no significant change in the results of paper and impacts on the metalicity and sSFR median data points by less than $\pm$0.01 and $\pm$0.2 dex, respectively.
	
	We note that the fluxes for each object are aperture - matched and deblended, and variations of PSF and pixel scale across the various bands are correctly accounted for, by preforming aperture photometry using the \textsc{LAMBDAR} software package \citep[see detail description of the algorithm in section 3 of ][]{Wright2016}.
	All fluxes in the GAMA samples from FUV to K bands are corrected for galactic extinction \citep{Wright2016}. For a morphological classification of our sample, we use the elliptical probability catalogue of GAMA. 
	This is based on the fraction of Galaxy Zoo votes for ellipticals \citep{Lintott2011}, combined with the ELLIPTICAL morphological classifications performed on postage stamps images from the SDSS and VIKING data as described in \citet{Driver2012} for the GAMA II sample.
	However, only half of our sample was classified in this catalogue. Therefore, we also used the catalogue of \cite{Kuminski2016} for the rest of the sample. In the second catologue, machine-learning was used to classify the broad morphological types of ~3,000,000 SDSS DR8 galaxies with a statistical agreement rate of ~98\% with the Galaxy Zoo debiased 'superclean' dataset. By combining these two catalogues, 93\% of our sample has a morphological classification(Table \ref{tab:value}). We confirm that the two catalogues provide comparable results, and thus can be safely combined. To do this, we compare the morphological classification of those objects which appear in both catalogues. This overlap sample is 20\% of the total sample. We find that 95\% of the overlapping sample agrees with the fractional probability of being an elliptical to within $\pm$0.2. Thus there is good agreement between the two morphological catalogues.
	
	\section{Analysis and results}
	\subsection{$\Delta M_{12} -D_{offset}$ relation }\label{Sec:Delta_D_off}
	Figure \ref{fig:lum_gap_NUV} shows the distribution of the magnitude gap, $\Delta M_{12}$ as function of BGG offset from the group's luminosity center, $D_{offset}$. Data points are color-coded by their NUV-r color. Black dashed lines show the regions for halos of different dynamical states from (1) to (6). Region (1) is relaxed while region (6) is unrelaxed, and these represent the two extremes of our sample in terms of dynamical state. The percentage in the red and blue areas of the pie-chart represents the percentage of galaxies with a color of NUV-r $>$4.5 (red) and $<$ 4.5 (blue) for each region. As can be seen in the region (1), the relaxed galaxy groups, we find more than 63$\pm$7\% of their BGGs are red galaxies. In contrast, we also find that more than 61$\pm$4\% percentage of BGGs in the region (6), the unrelaxed systems, are blue galaxies.
	As can be seen in the bottom panels of the figure, both the offset and luminosity gap of the groups affect the NUV-r colors of their BGGs. With increasing luminosity gap, there is an increasing red fraction, and similarly with decreasing BGG offset. However, it can be seen that the colour fraction is most sensitive to the luminosity gap.
	
	We note that, throughout this paper, we compare several different subsamples -- relaxed/unrelaxed, large/small gap and large/small offset. We expect that the relaxed/unrelaxed sample will be the most extreme comparison. However, the reason for considering the gap and offset individually is two-fold. Firstly, it allows us to clearly see the efficiency of the gap and offset parameter separately on the results. Secondly, it enables us to test our results are robust to low number statistics. For example, the sample size is reduced by 30\% (see Table \ref{tab:value}) if we use the gap and offset parameters combined, which results in an increase in the Poisson errors (shown above pie-charts).
	
	\subsection{Dependency on morphology}
	As can be seen in the Figure \ref{fig:pellipgama}, the BGGs in unrelaxed groups have a higher fraction of non-elliptical (disk) galaxies with respect to the relaxed groups with the luminosity gap being the most important driving parameter behind the differences seen between the BGGs, more so than the BGG offset parameter (median and mean elliptical probability report in Table \ref{tab:value}). In light of this, there appears to be a correlation between the dynamical state of galaxy groups and the morphological type of their BGGs, perhaps through their formation history, although the normal distributions are similar within a one-sigma error (see SD error in table \ref{tab:value}). This likely contributes to the fact that the unrelaxed groups have bluer BGGs but, as we will show now, this is not the entire story. 
	
	In Figure \ref{fig:NUV_flux_fossil_ctrl_scater}, we plot the distribution of {\rm NUV-r} as function of BGG's sersic index in the r-band for the relaxed and unrelaxed samples. The red line and blue dashed-line show the median binned data points for the BGGs in relaxed and unrelaxed galaxy groups, and it is clear that NUV-r colors are typically bluer in unrelaxed systems, even at the same sersic index. If we take sersic index as a rough proxy for galaxy morphology, this suggests that the bluer NUV-r colors of the unrelaxed group BGCs are not just because there are more disky galaxies in that sample, but also that the stellar populations are genuinely bluer, even at fixed morphology.
	\begin{figure*}
		\includegraphics[width=0.98\textwidth]{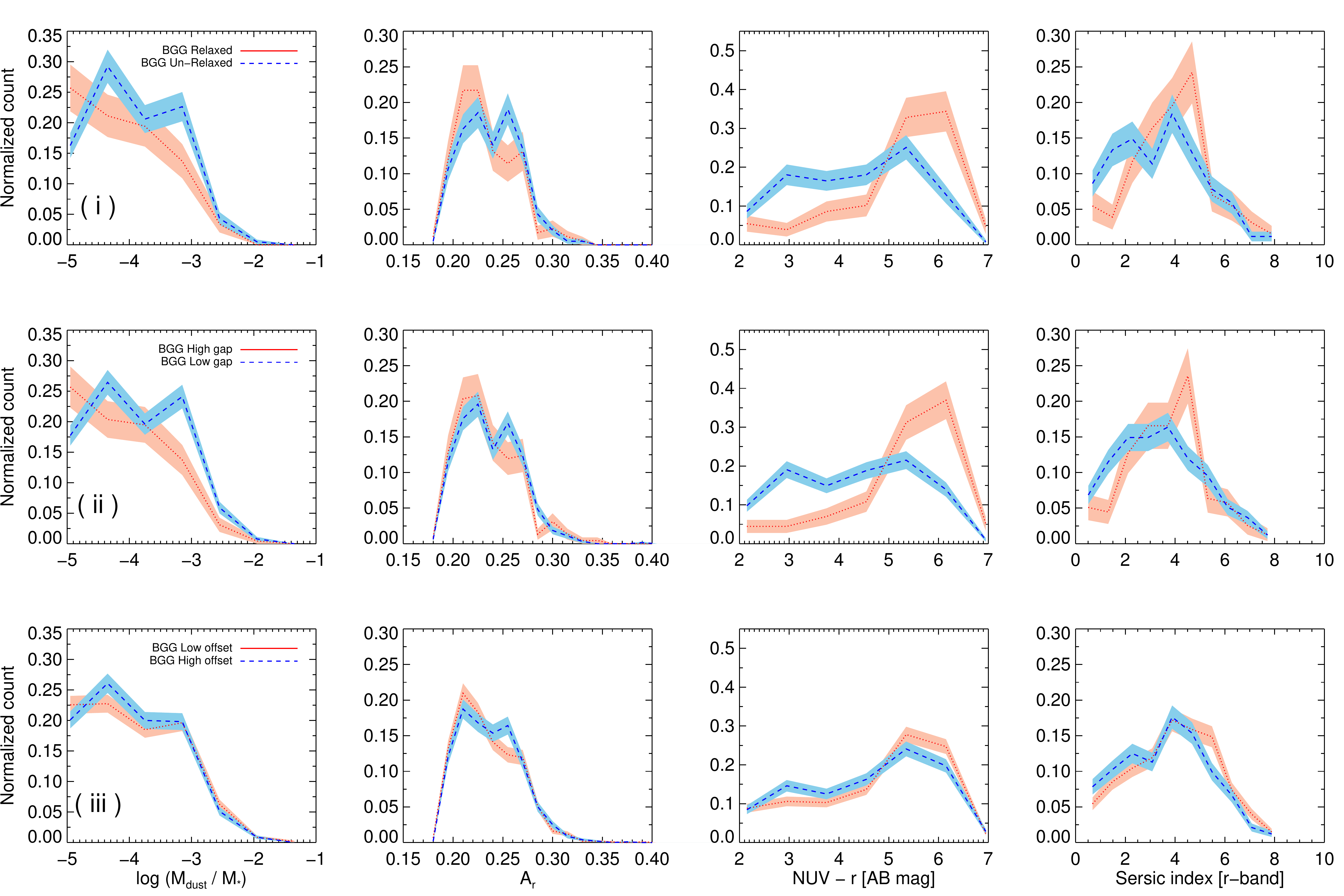}
		\caption{Histograms of specific dust mass, galactic extinction in r-band ($A_r$), NUV - r and sersic index for different categories of groups selected on the basis of (i) relaxed/unrelaxed  (top panels), (ii)high/low -- gap, $\Delta M_{12}$, (middle panels), and (iii) low/high -- offset, $D_{offset}$, (bottom panels) illustrated by red-lines/blue-dashed-line and red-shade/skyblue-shade as Poisson error bars.}
		\label{fig:gap_distribution}
	\end{figure*}
	
\begin{figure}
		\centering
		\includegraphics[width=0.4\linewidth]{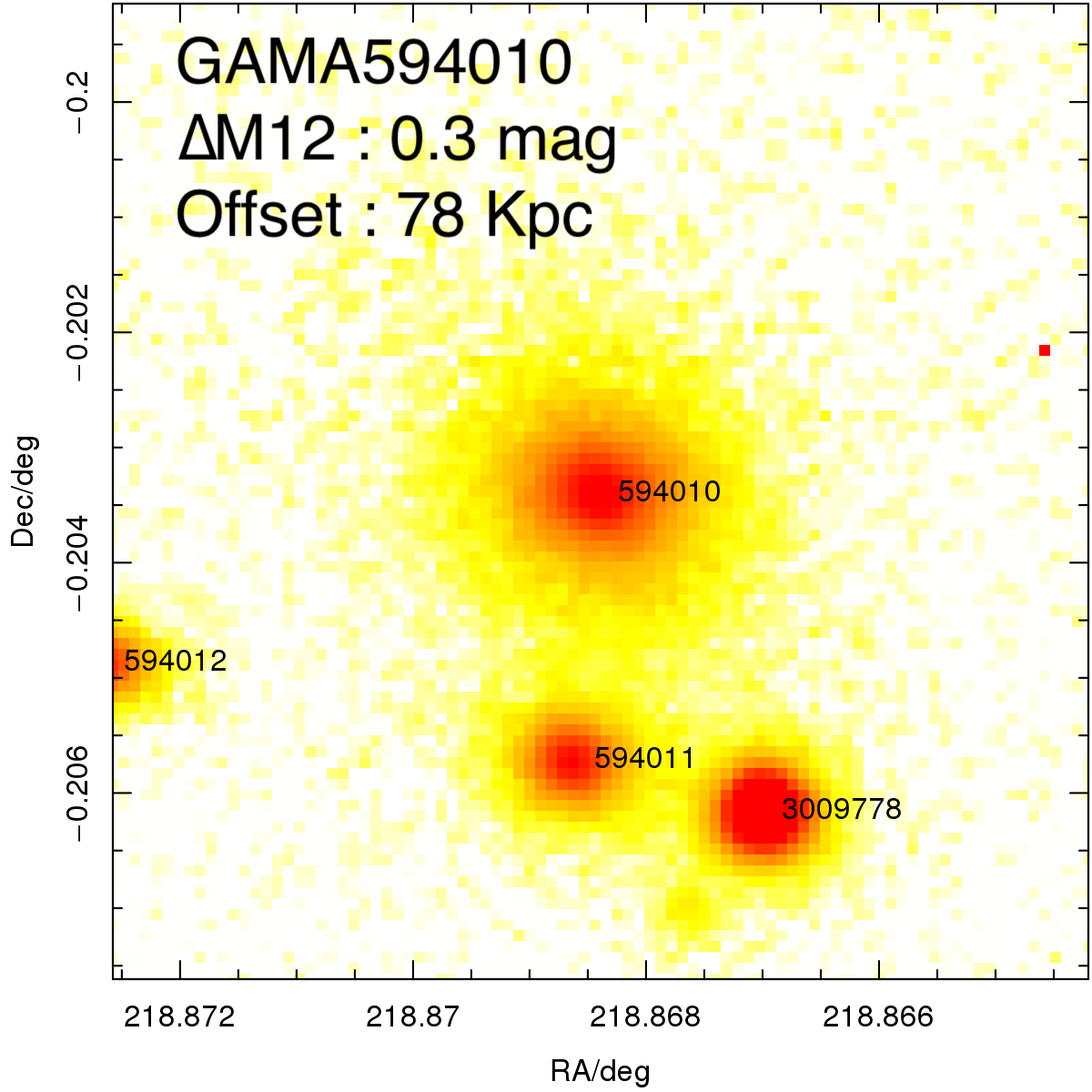}
		\includegraphics[width=0.4\linewidth]{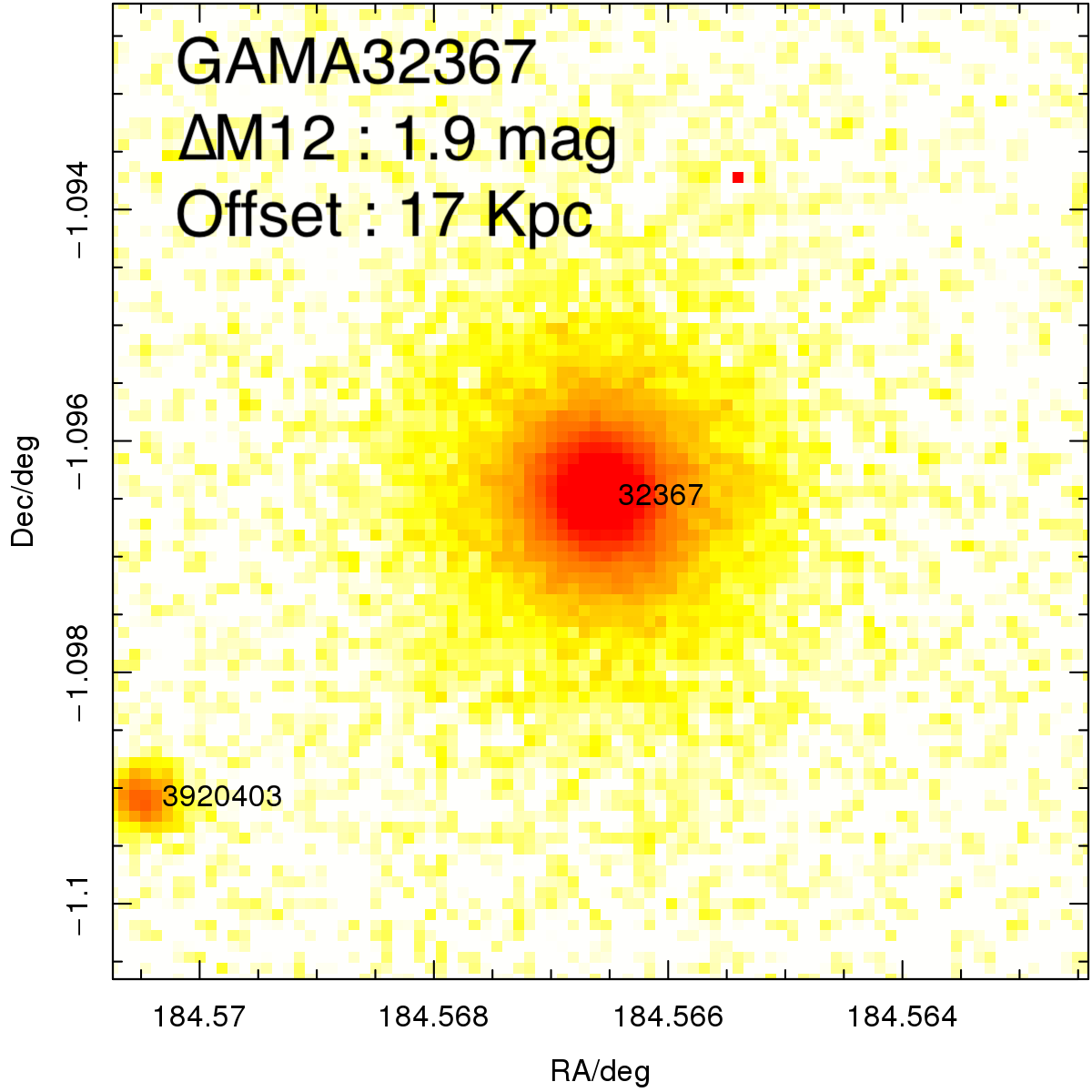}
		\includegraphics[width=0.4\linewidth]{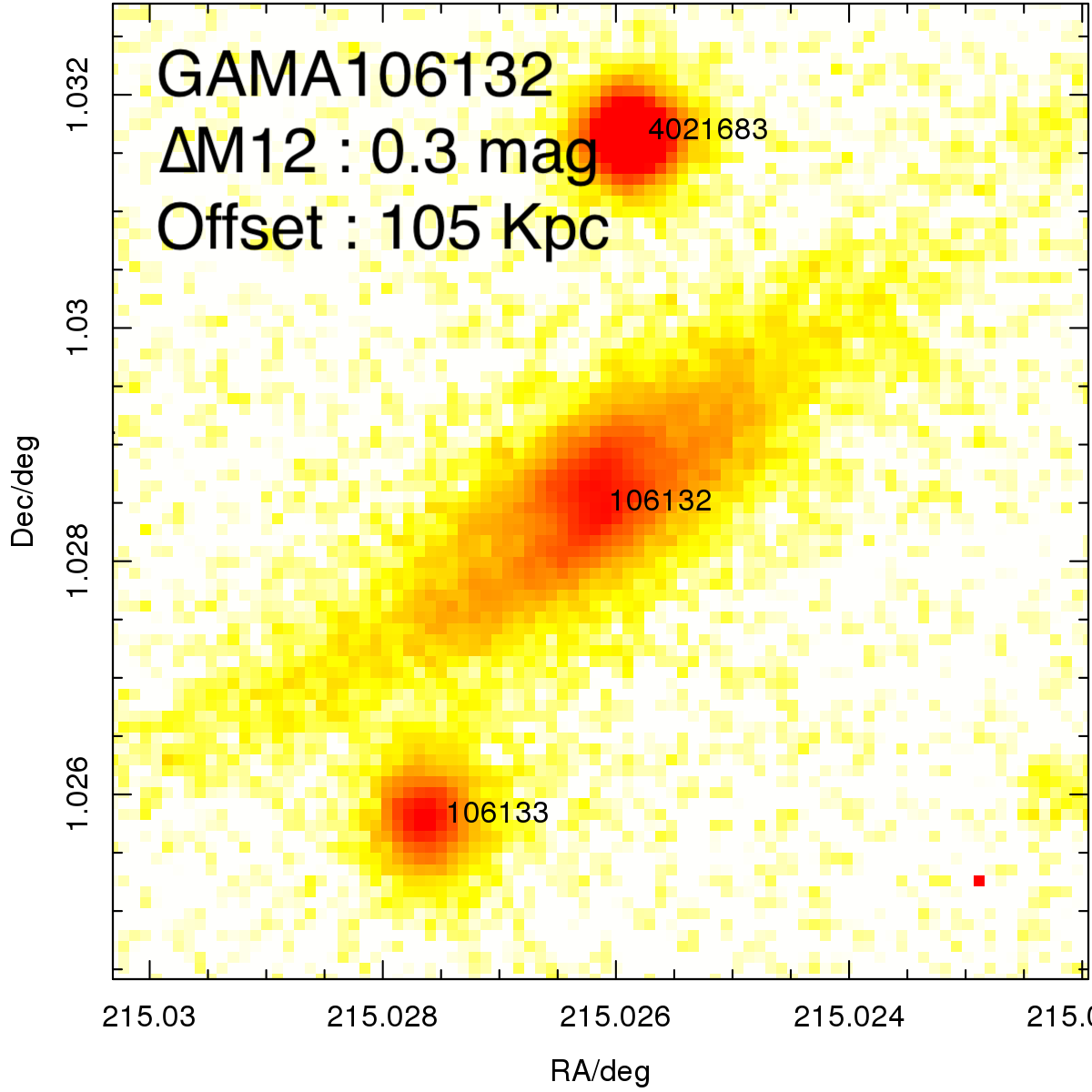}
		\includegraphics[width=0.4\linewidth]{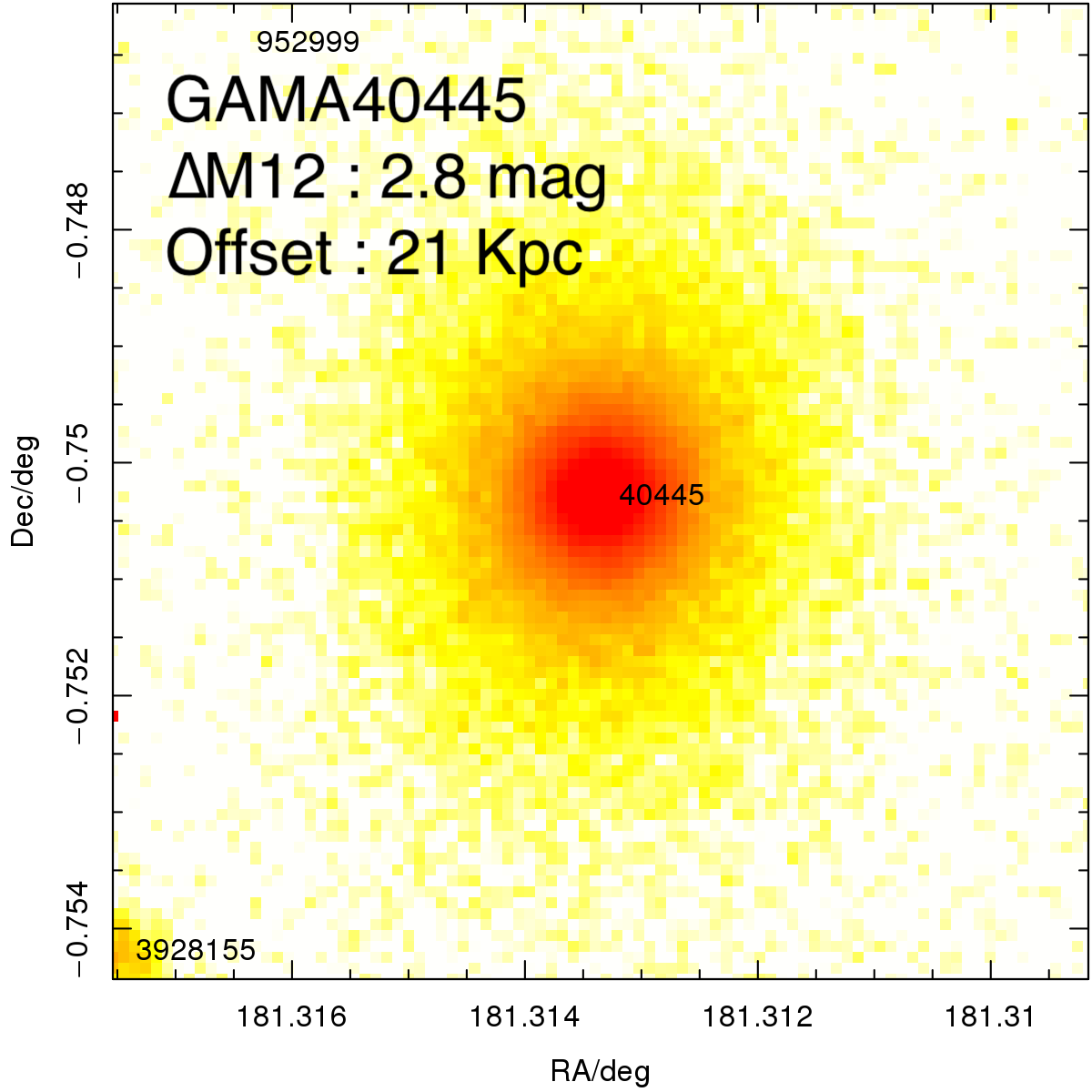}
		\includegraphics[width=0.4\linewidth]{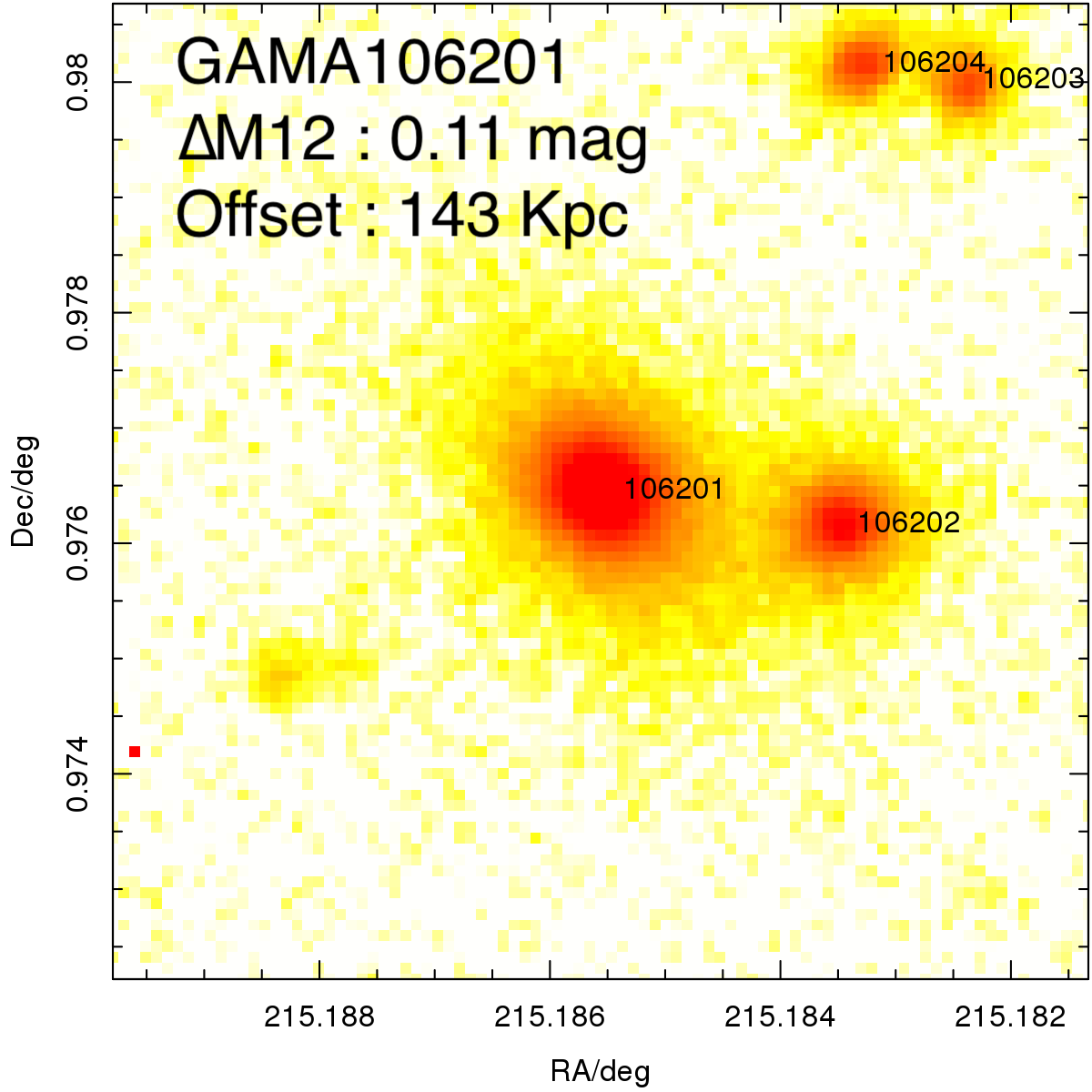}
		\includegraphics[width=0.4\linewidth]{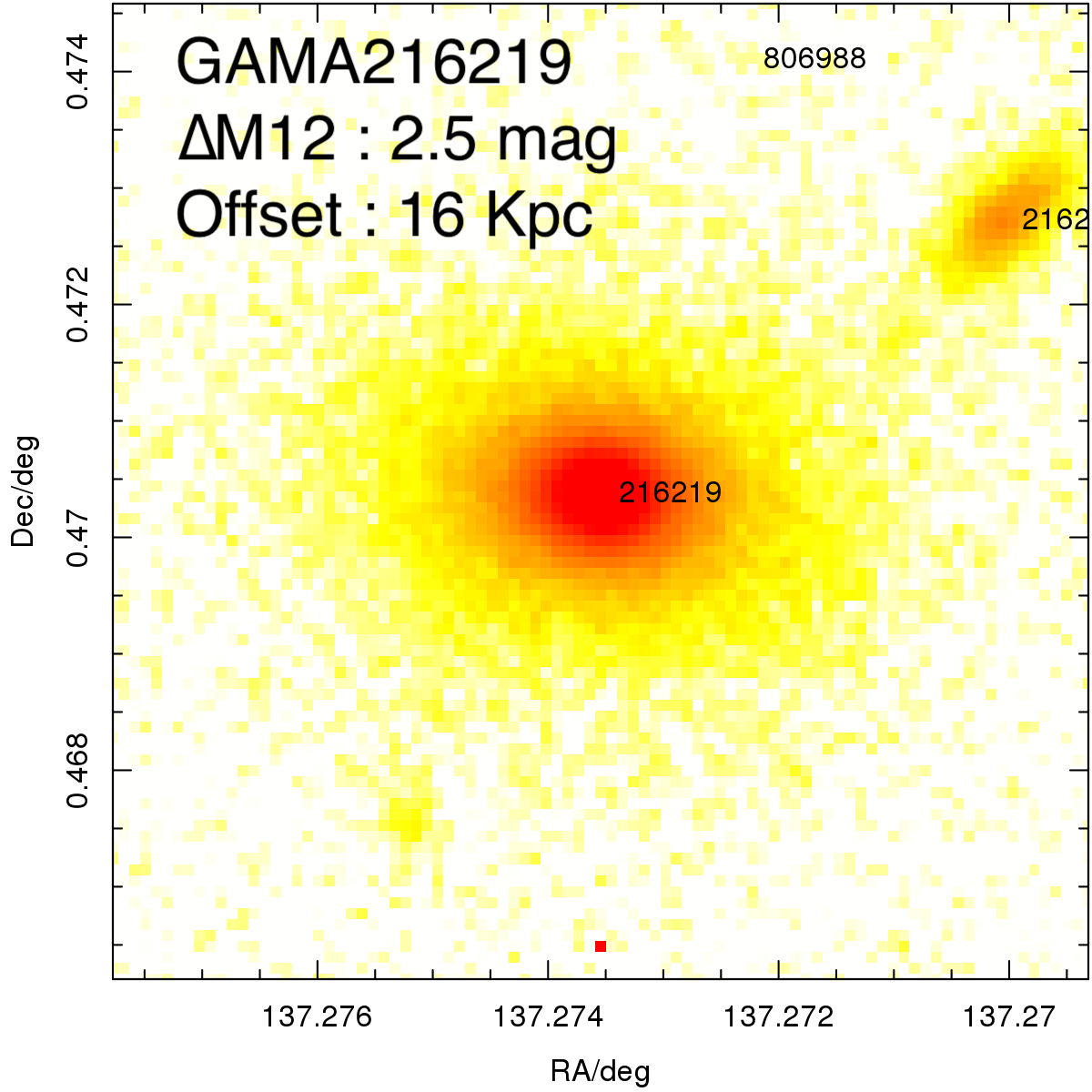}
		\includegraphics[width=0.4\linewidth]{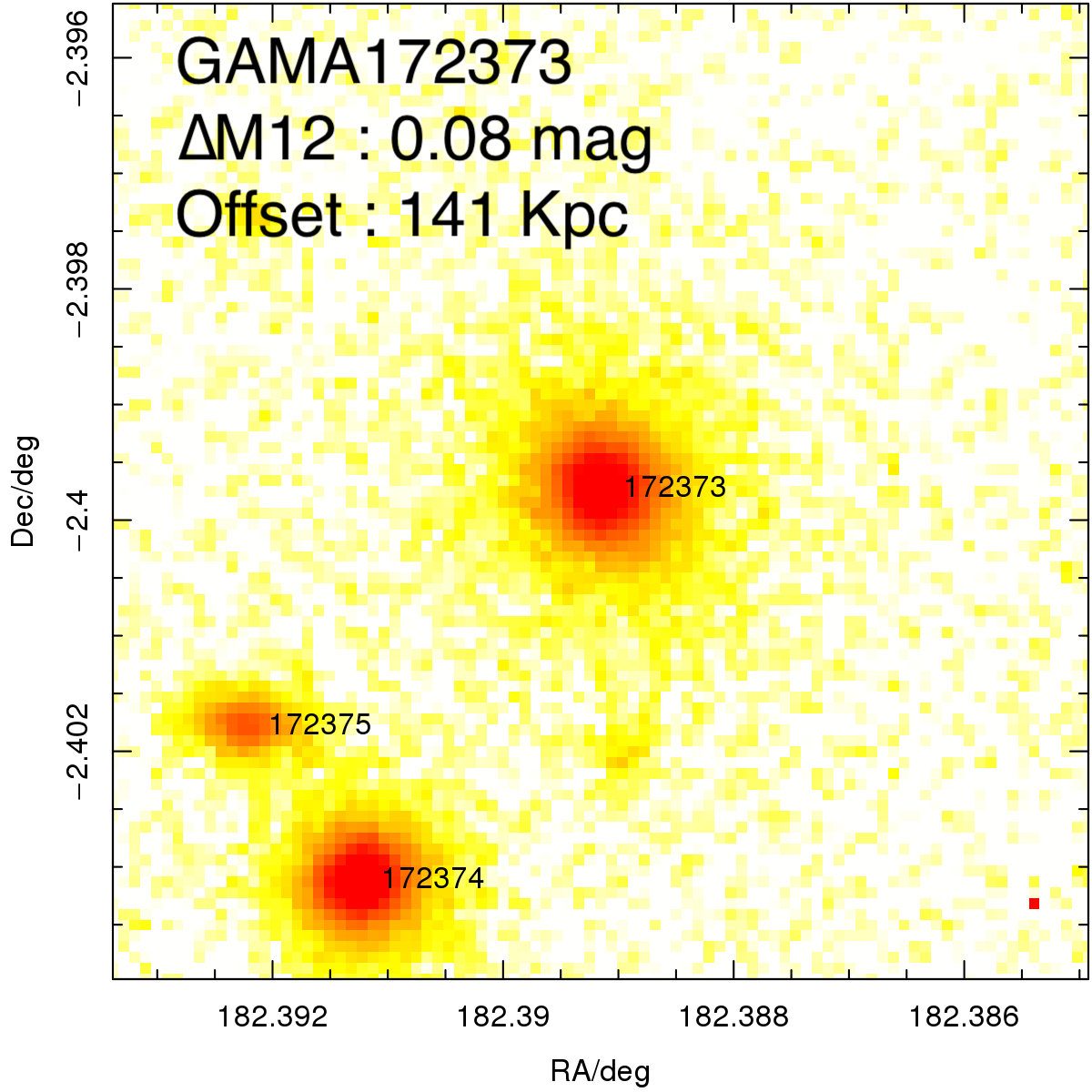}
		\includegraphics[width=0.4\linewidth]{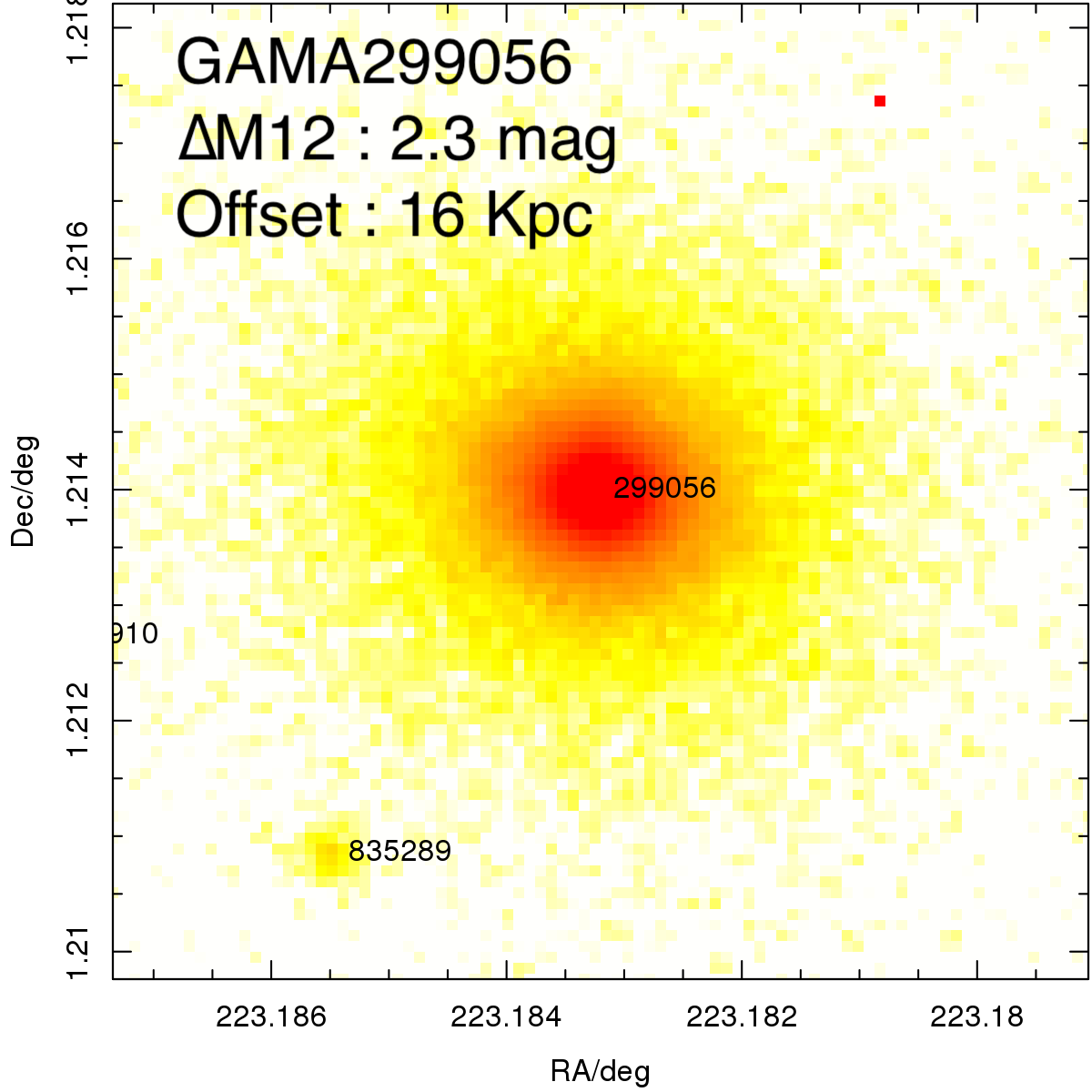}
		\includegraphics[width=0.4\linewidth]{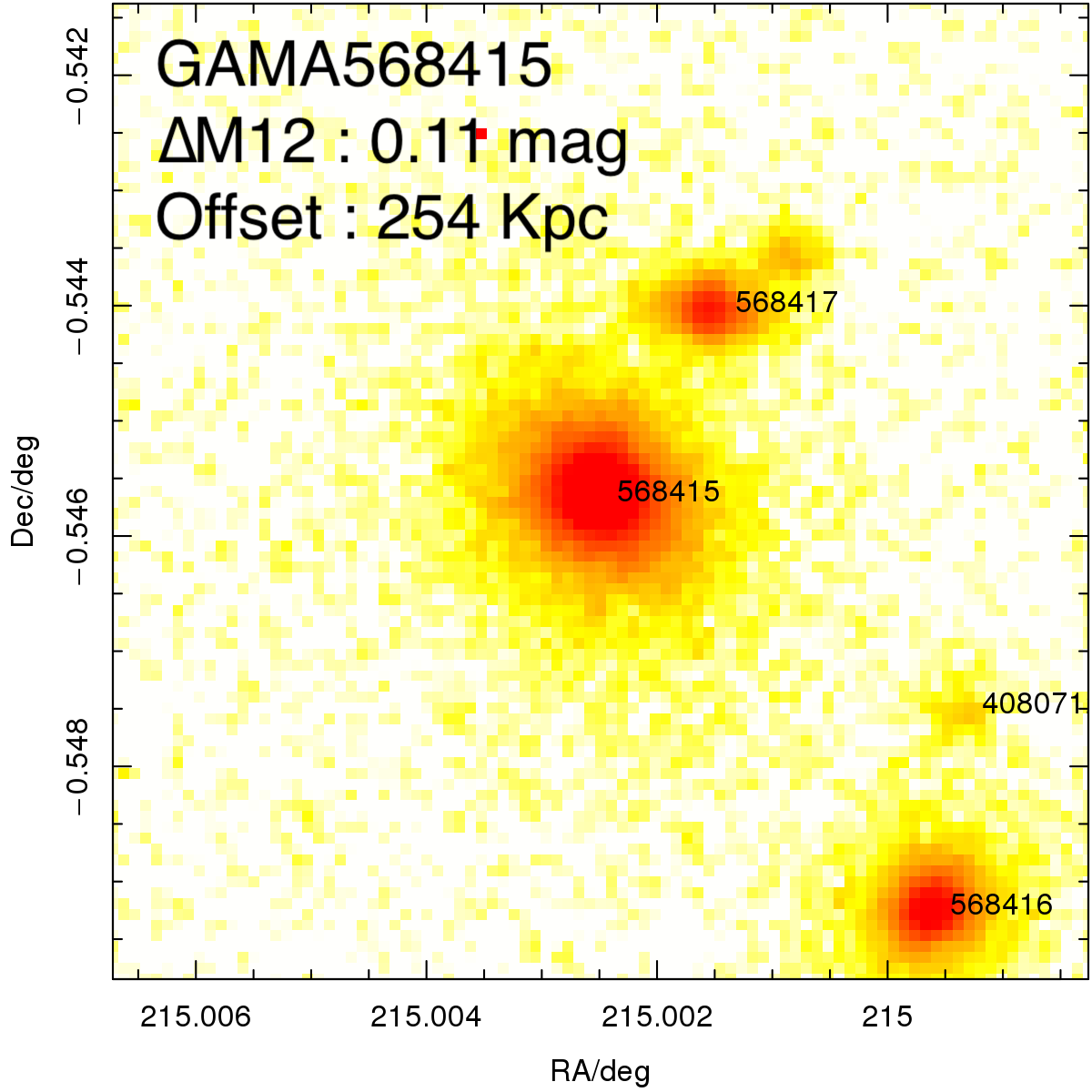}
		\includegraphics[width=0.4\linewidth]{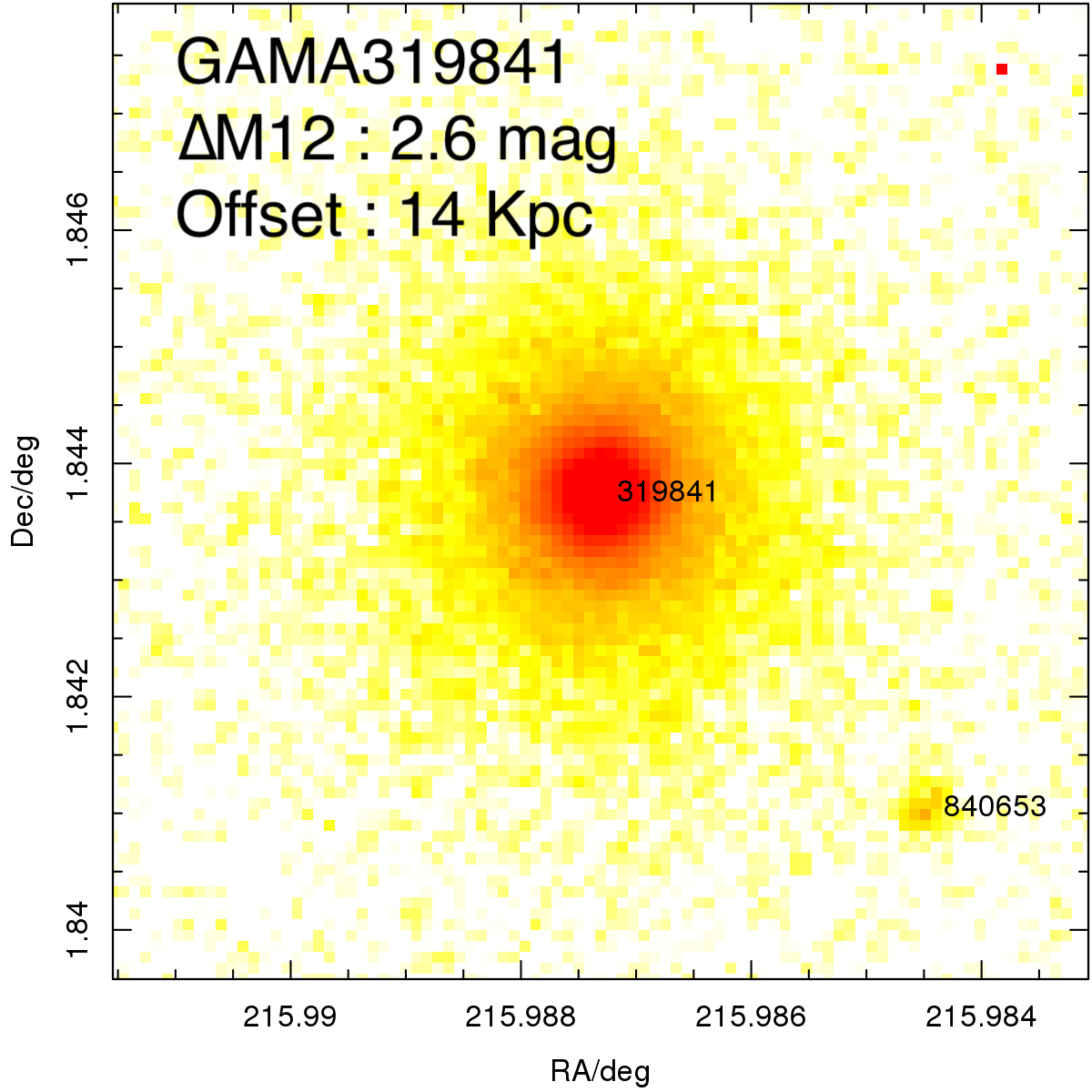}
		\includegraphics[width=0.4\linewidth]{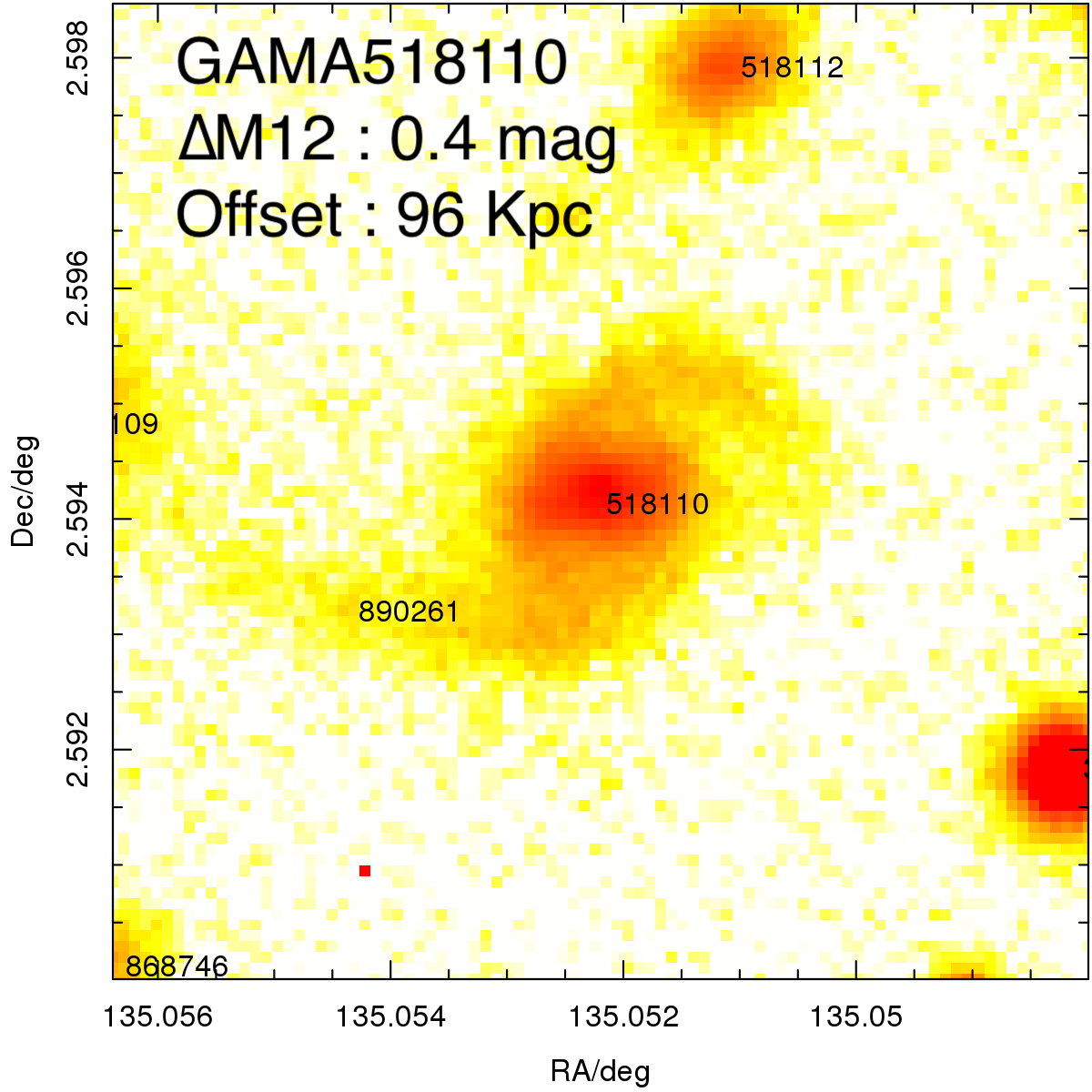}
		\includegraphics[width=0.4\linewidth]{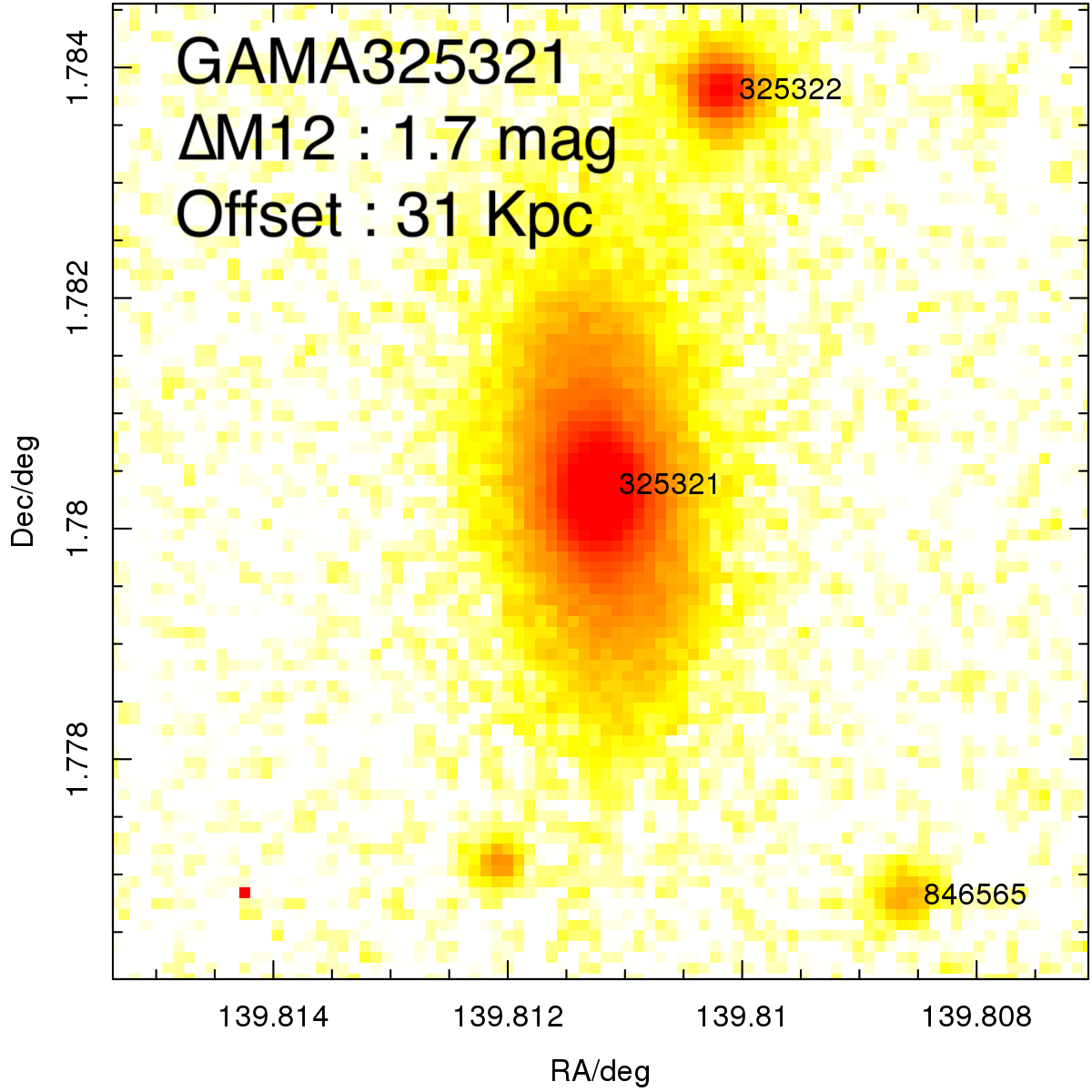}
		
		\caption{Selected samples of 6 unrelaxed (left panels) and 6 relaxed (right panels) representative samples of an optical r-band inverted image extracted from Sloan Digital Sky Survey using the GAMA Panchromatic Swarp Imager (PSI). It can be seen here and in Figure \ref{fig:pellipgama} that there are more disky BGGs in the unrelaxed sample (e.g. the median probability of being an elliptical is only 68\% in unrelaxed groups compared to 82\% in relaxed groups).  All images are 15" cutouts, and the fields are visibly more crowded in the unrelaxed sample. We also report the GAMA name, $\Delta M_{12}$ and BGG offset in each selected sample. }
		\label{fig:Unrelax}
	\end{figure}

	Figure \ref{fig:gap_distribution} shows from left to right histograms of the specific dust mass, galactic extinction ($A_r$), NUV - r and sersic index for different categories of groups from top to bottom selected on the basis of (i) relaxed/unrelaxed, (ii) high/low -- gap $\Delta M_{12}$, and (iii) low/high -- offset $D_{offset}$ . The subpanels of the above figure show the effects of the magnitude gap ($\Delta M_{12}$) in the bimodality distribution of NUV-r color and specific dust mass, while there is a less of a significant difference between the low/high -- offset sub-samples, similar to the results in Figure \ref{fig:lum_gap_NUV}.
	As can be seen in the figure, while the relaxed peaks are typically more de Vaucouleurs in the form (sersic index = 4), the unrelaxed group sersic index distribution peaks at lower values, and may contain an additional contaminating sample of more disk-like objects with sersic index$\sim$2. We also see that the dust mass and galactic extinction of BGGs in unrelaxed groups is higher than that of the relaxed groups and the difference is once again mostly driven by the group luminosity gap.
	The r-band images of a small representative sample of relaxed and unrelaxed BGGs are shown in Figure \ref{fig:Unrelax}. The figure demonstrates the more crowded fields and the higher frequency of disk-like morphologies in the BGGs of the unrelaxed sample with respect to the relaxed samples. Note that we select the representative sample randomly from the subsample of relaxed and unrelaxed groups, and we visually confirm that the morphologies of the representative sample approximately match the expected morphological fractions found in Figure \ref{fig:pellipgama}. 
	
	\subsection{Dust mass counterparts}
\begin{figure}
		\includegraphics[width=0.5\textwidth]{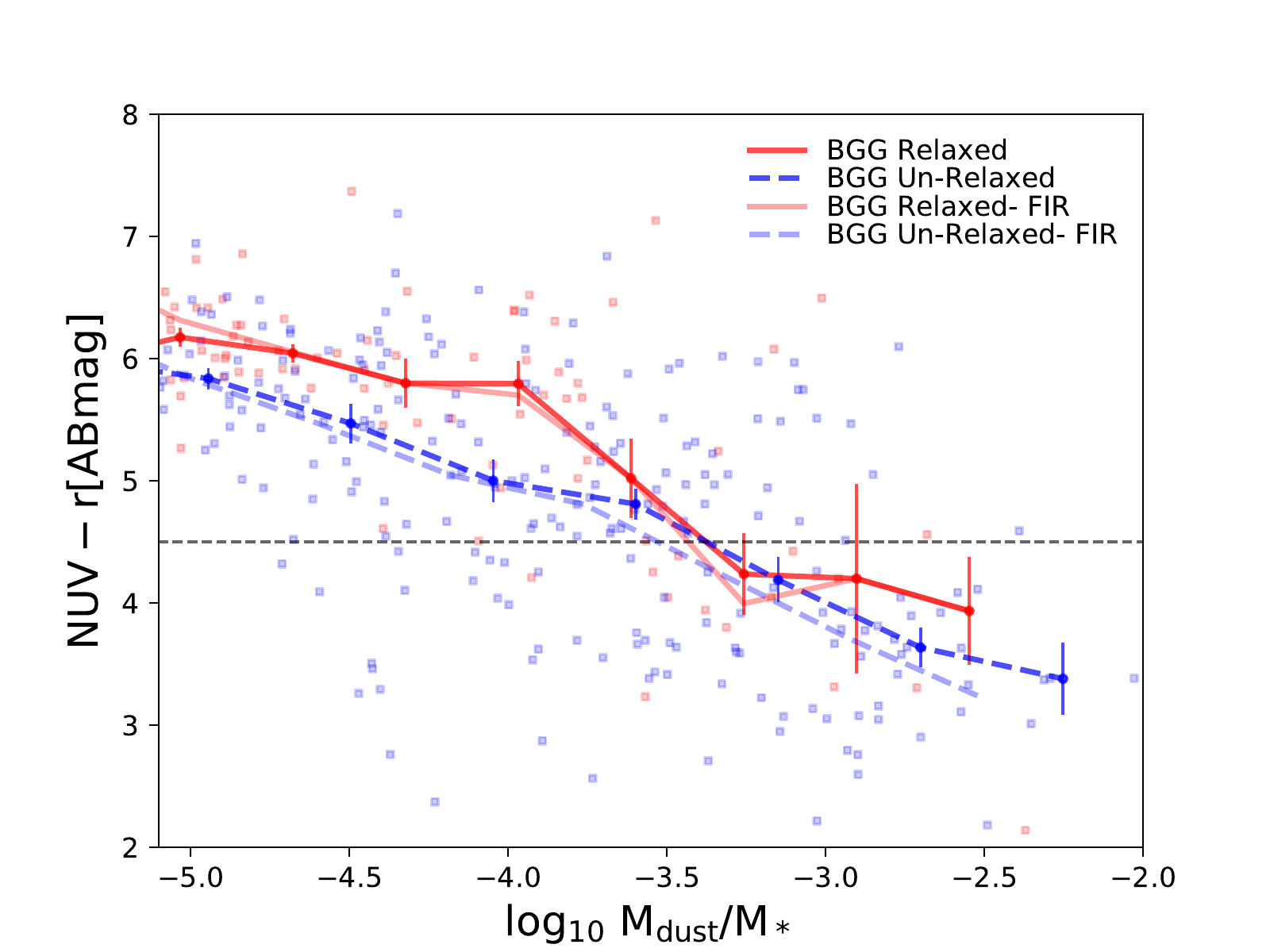}
		\caption{Distribution of NUV-r color as a function of specific dust mass for relaxed and unrelaxed galaxy groups.
			The \emph{red line} and \emph{blue dashed-line} show the medians and $\sigma/\sqrt{N}$  uncertainties for BGGs of relaxed and
			unrelaxed groups, respectively. Faint color lines show the same trend for the sub-samples with excluding the undetected FIR data.
			The horizontal dotted-line shows the NUV-r=4.5 boundary used to divide the sample into red and blue galaxies. Note that at $log_{10}\ (M_{dust}/M_*) >$ -3.5 the BGGs in relaxed and unrelaxed are indistinguishable in the NUV-r color within the error.}
		\label{fig:Dust_Mass_NUV_r}
	\end{figure}

	Dust absorption can have a significant impact on the total spectral energy distribution (SED) of a galaxy: absorbed stellar light in the UV and visible wavelengths is re-emitted in the IR by the dust grains. Therefore in this section, we try to test if the differences we observe in NUV-r color between relaxed or unrelaxed groups remain at fixed dust mass, as shown in Figure \ref{fig:Dust_Mass_NUV_r}. As can be seen in the figure, the BGGs of relaxed groups tend to be redder in NUV - r color at a given specific dust mass. 86\% of BGGs in relaxed groups have NUV - r color $>$ 4.5 mag compared to 65\% of BGGs in unrelaxed groups. Note that the NUV-r color is indistinguishable  at $log_{10}\ (M_{dust}/M_*) > -3.5$ within the error.
	Note that we also find similar results if we instead use the $A_r$ extinction instead of dust mass, in terms of bluer colours for unrelaxed BGGs.
	Given that $NUV - r$ is sensitive to recent star formation, this implies that the majority of BGGs in relaxed groups are more passive. To further test this, in the following section we will directly compare the star formation rates derived from the SED fitting.
	The faint colored lines in Figure \ref{fig:Dust_Mass_NUV_r} show the results if we exclude galaxies that are undetected in the FIR from our sample. We see that our results are not significantly changed by their exclusion. 
	
	\begin{table*}
		\centering
		\large
		\caption{The total number count of groups in our sample (column 2) and the total number of objects with an available probability of being an elliptical galaxy, P(E), after combining the morphological classifications from GAMA and \citet{Kuminski2016}, the median, mean, standard deviation(SD) and the number of WISE cross-matched BGGs.}
		\label{tab:value}
		\begin{tabular}{lcccccc}
			\hline \\
			Samples         &   GAMA Count  & P(E) Count & Median(P(E)) & Mean(P(E)) & SD(P(E))  & WISE Count  \\
			\hline \hline
			All       &       1685     &  1544    &  0.79        &        0.66   &  0.14   &  1667 \\
			Relax     &       139       & 126  &  0.86          &       0.74  &  0.12    & 139 \\
			Un-Relax      &     399   &   352   & 0.73        &       0.6  &  0.15   & 392 \\
			High Gap      &     190       & 176 & 0.85        &         0.73  &  0.13  & 188\\
			Low Gap      &     598      & 532  &  0.73         &        0.60  &  0.15   & 591\\
			High offset      &     887      & 808 & 0.78         &        0.64  &  0.14    & 870\\
			Low offset      &      798     & 736 & 0.8          &        0.67   &  0.15   & 797 \\
			\hline \\
		\end{tabular}
	\end{table*}

	\begin{figure*}
		\centering
		\includegraphics[width=0.33\linewidth]{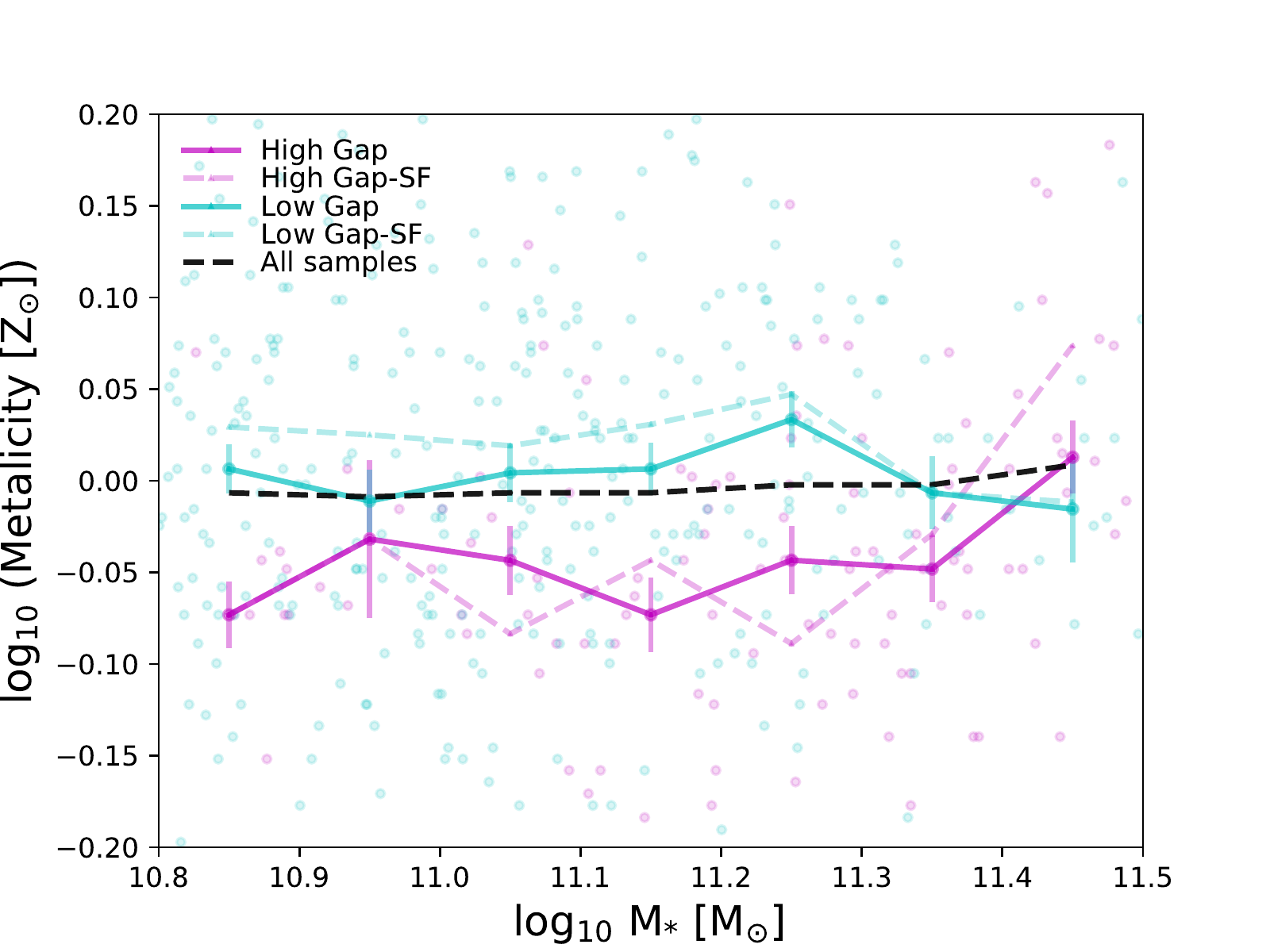}
		\includegraphics[width=0.33\linewidth]{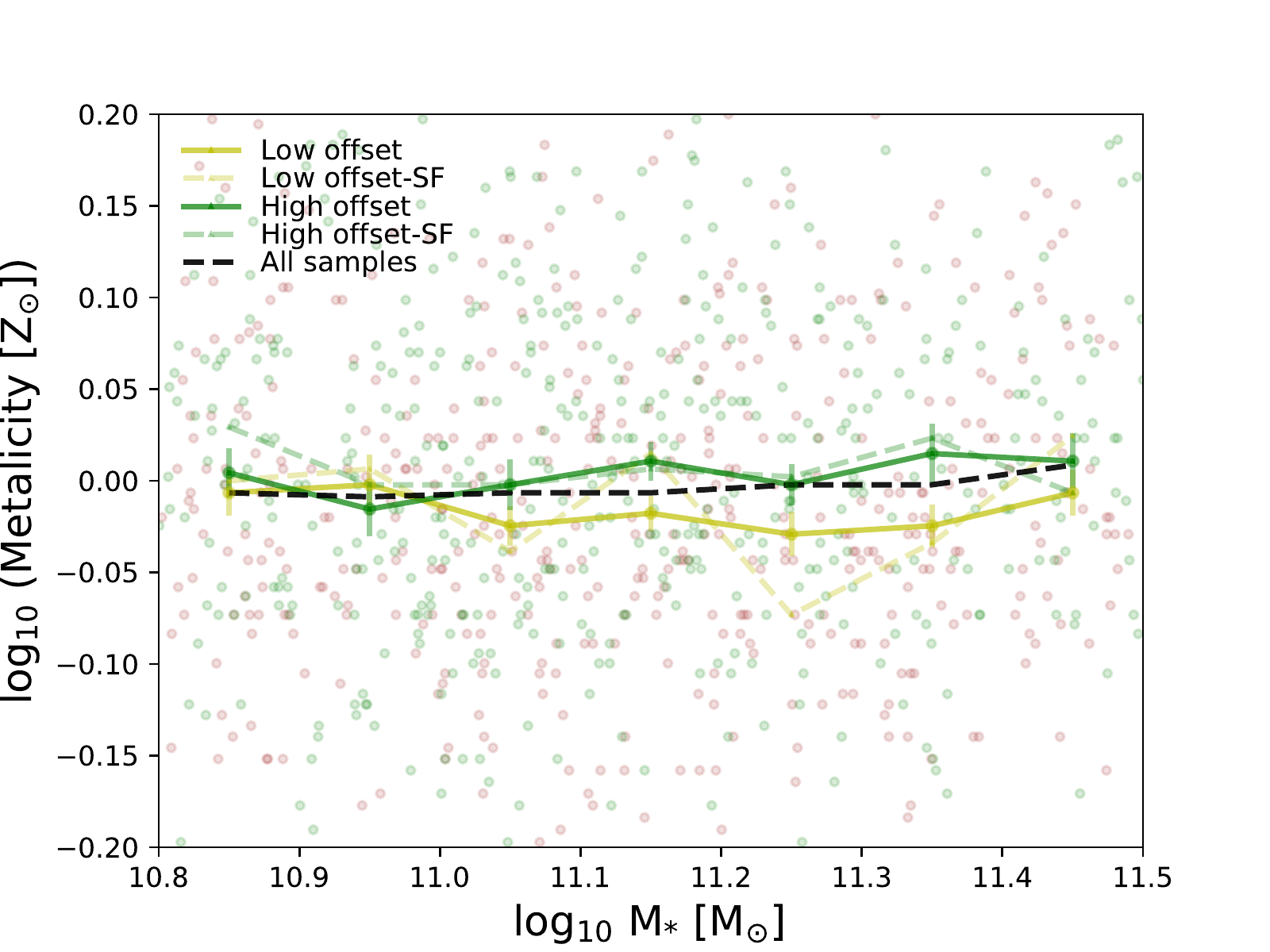}
		\includegraphics[width=0.33\linewidth]{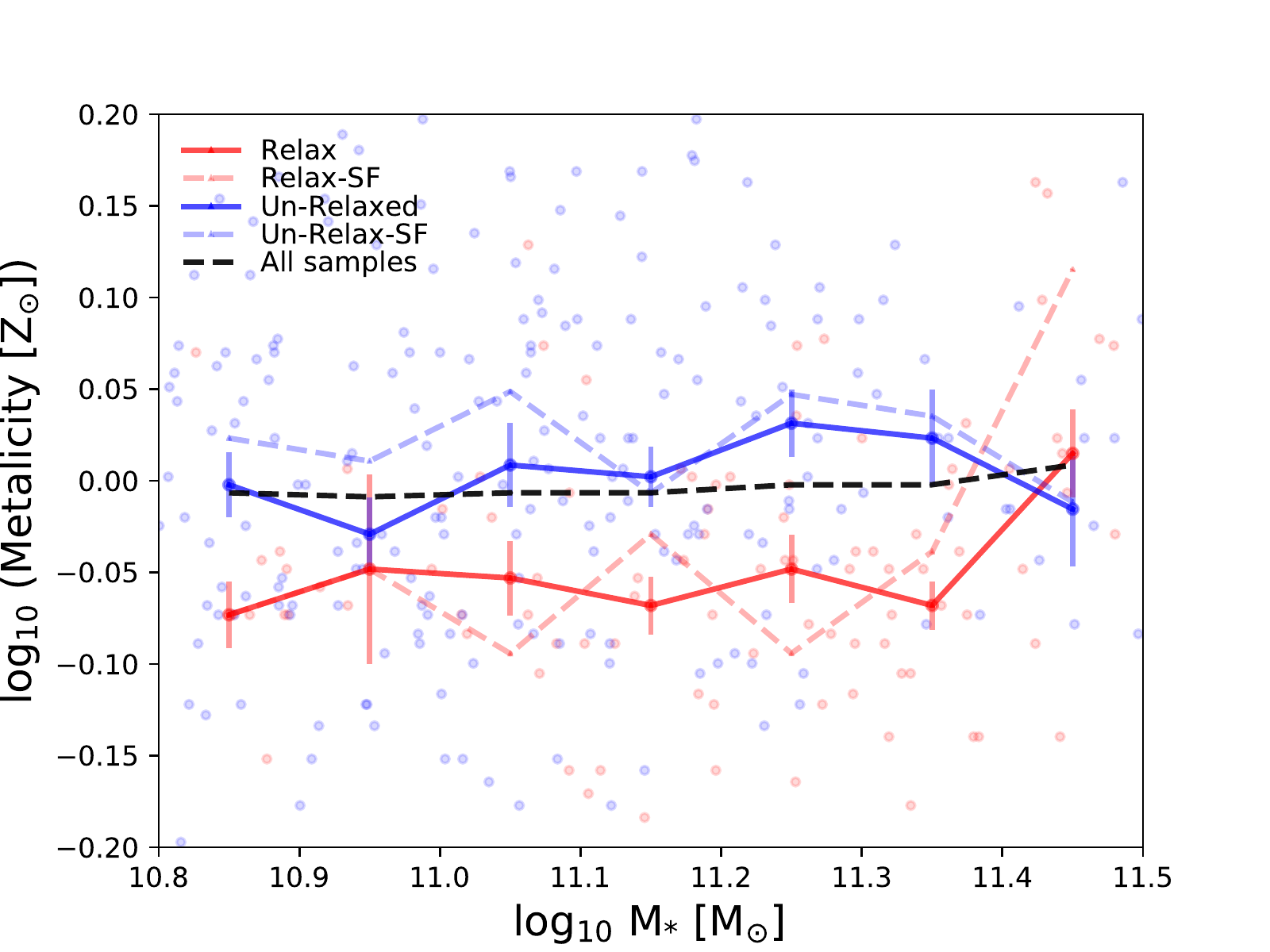}
		\includegraphics[width=0.33\linewidth]{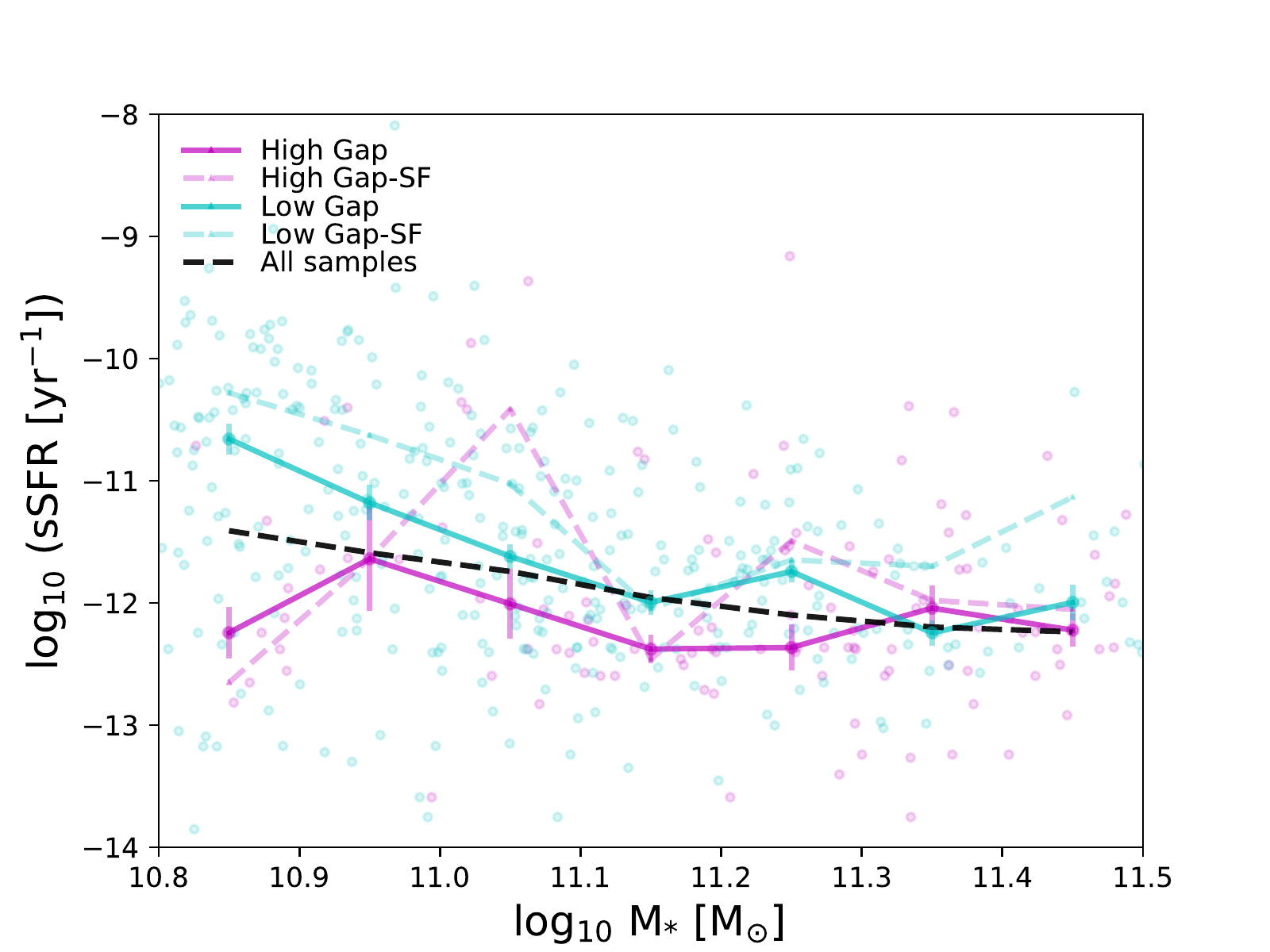}
		\includegraphics[width=0.33\linewidth]{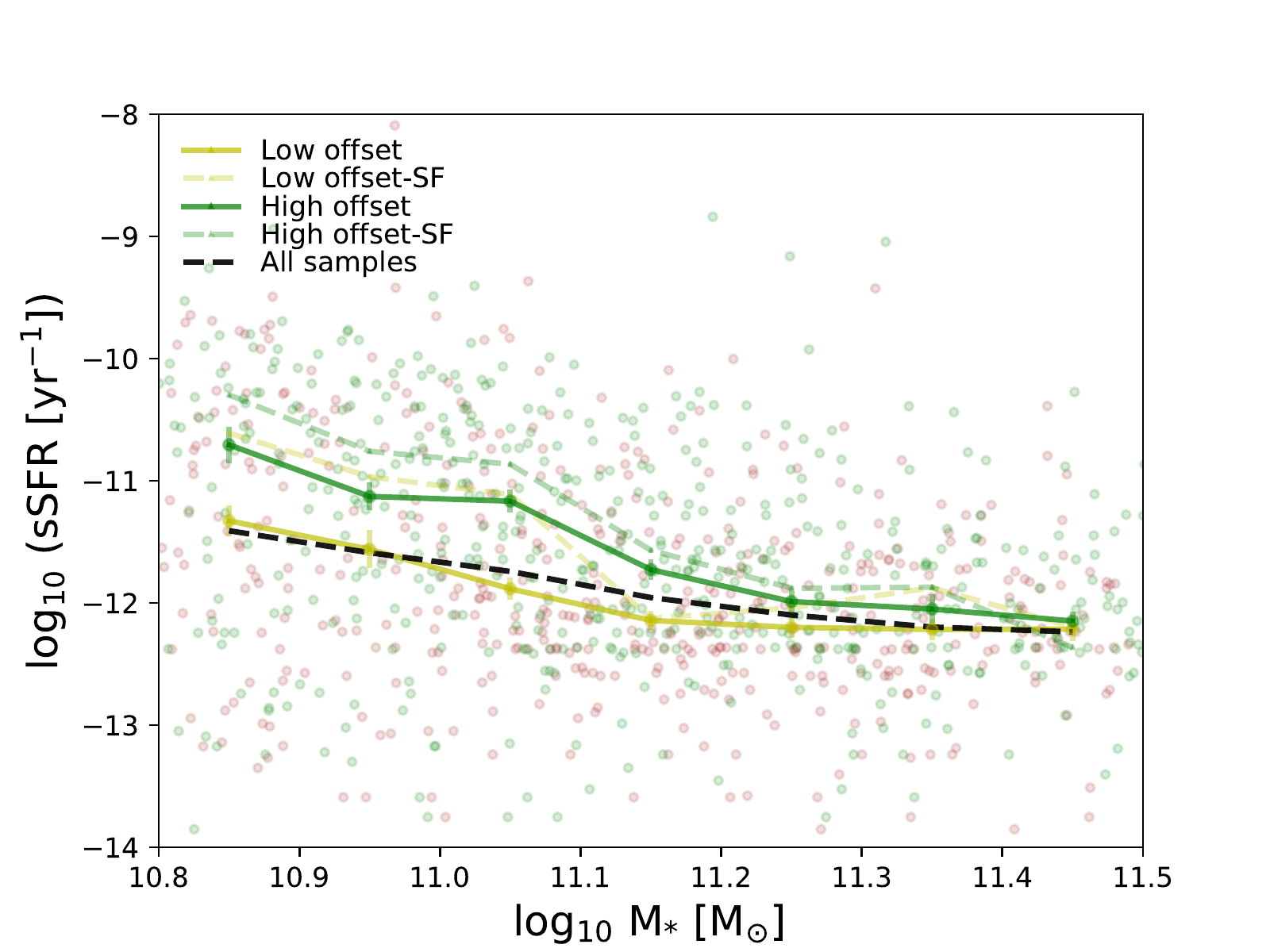}
		\includegraphics[width=0.33\linewidth]{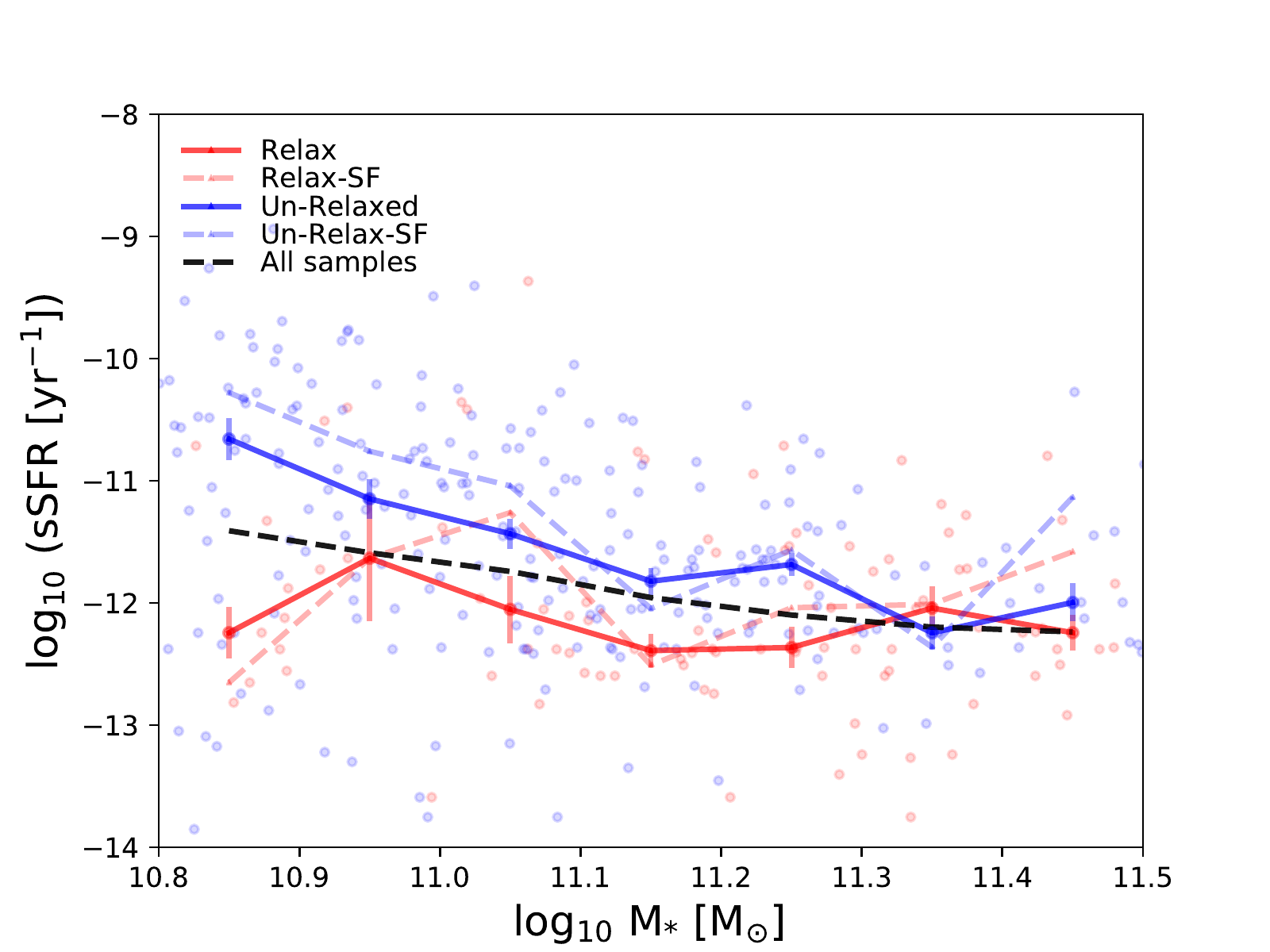}
		\includegraphics[width=0.33\linewidth]{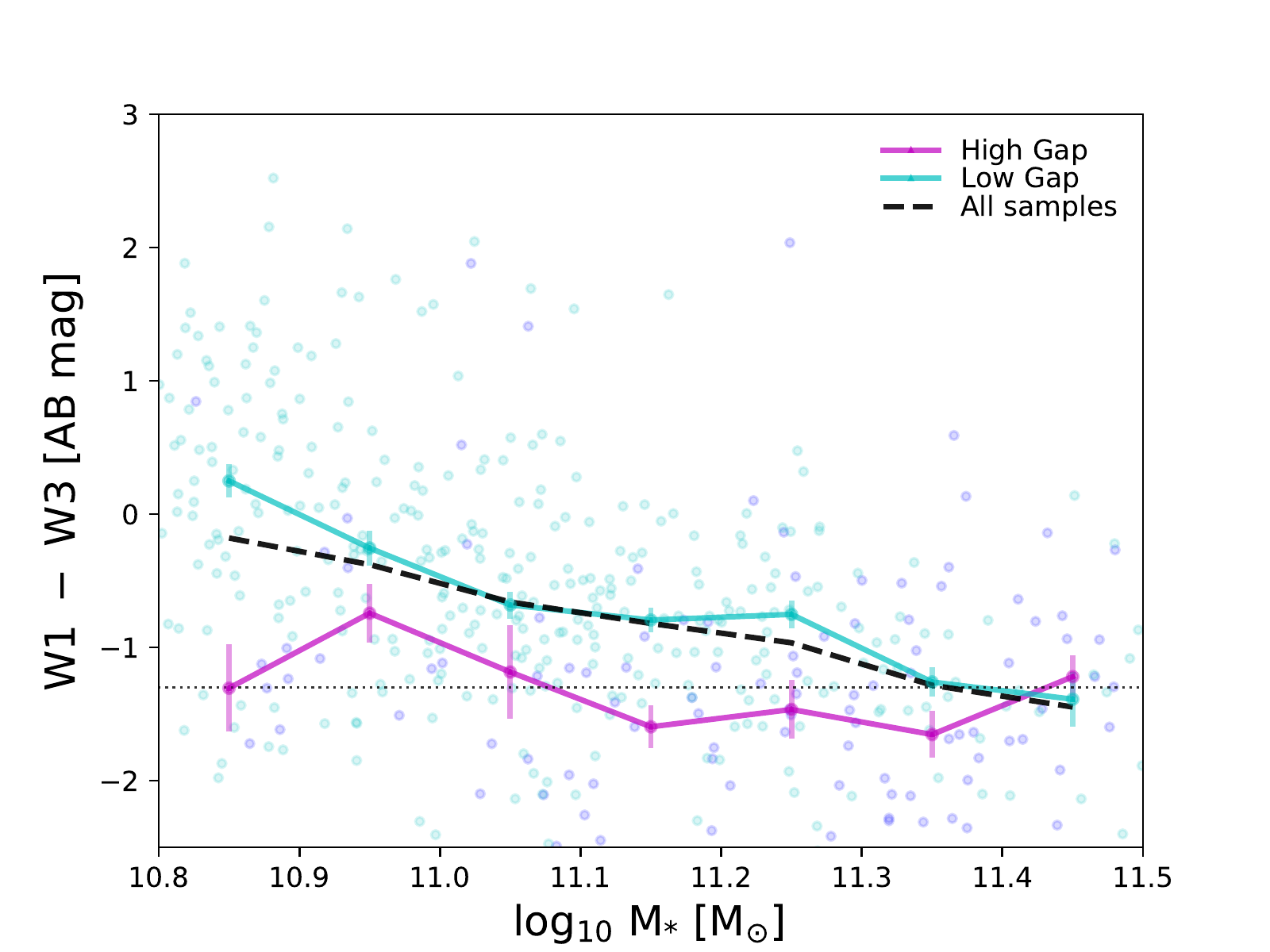}
		\includegraphics[width=0.33\linewidth]{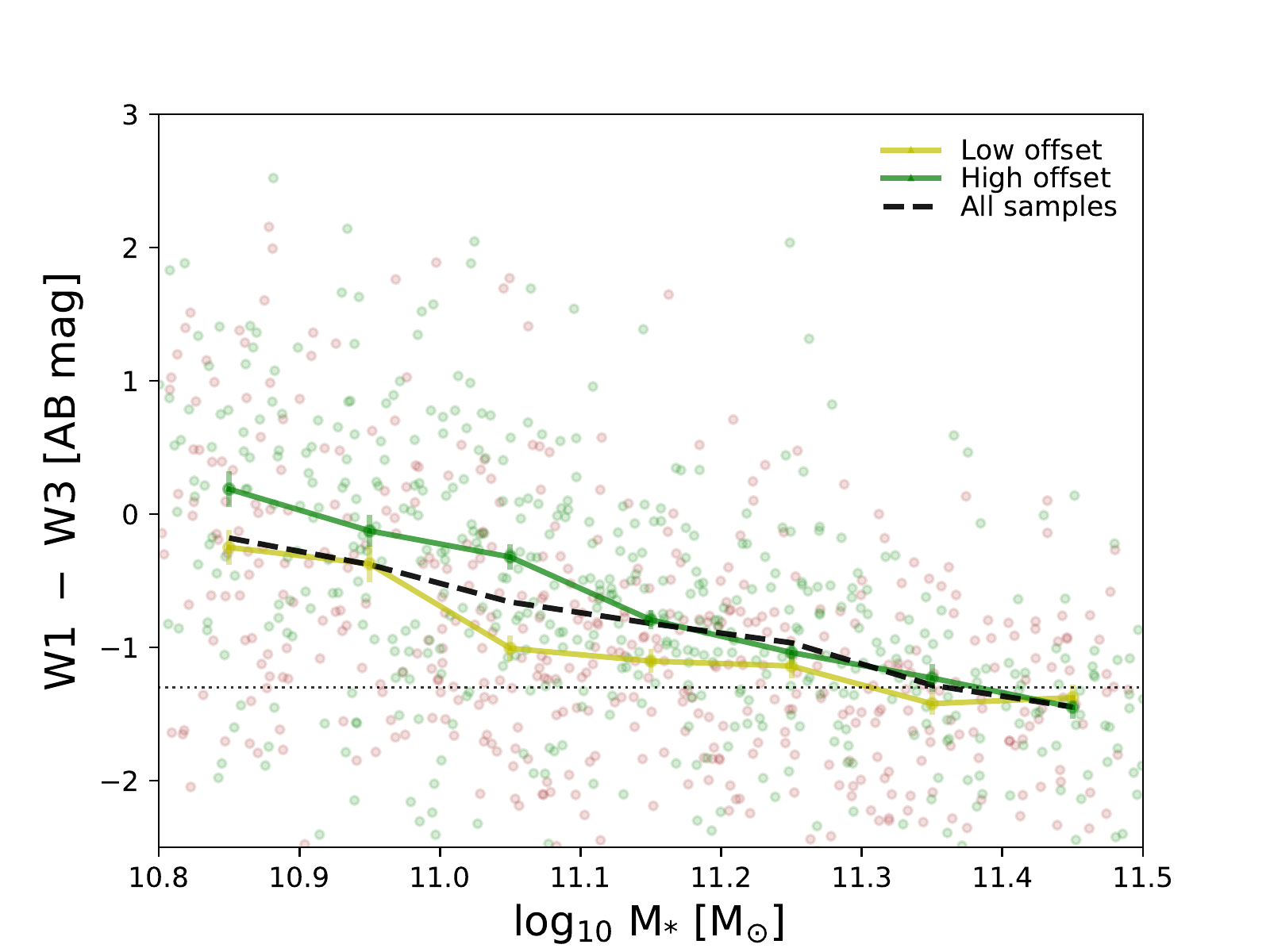}
		\includegraphics[width=0.33\linewidth]{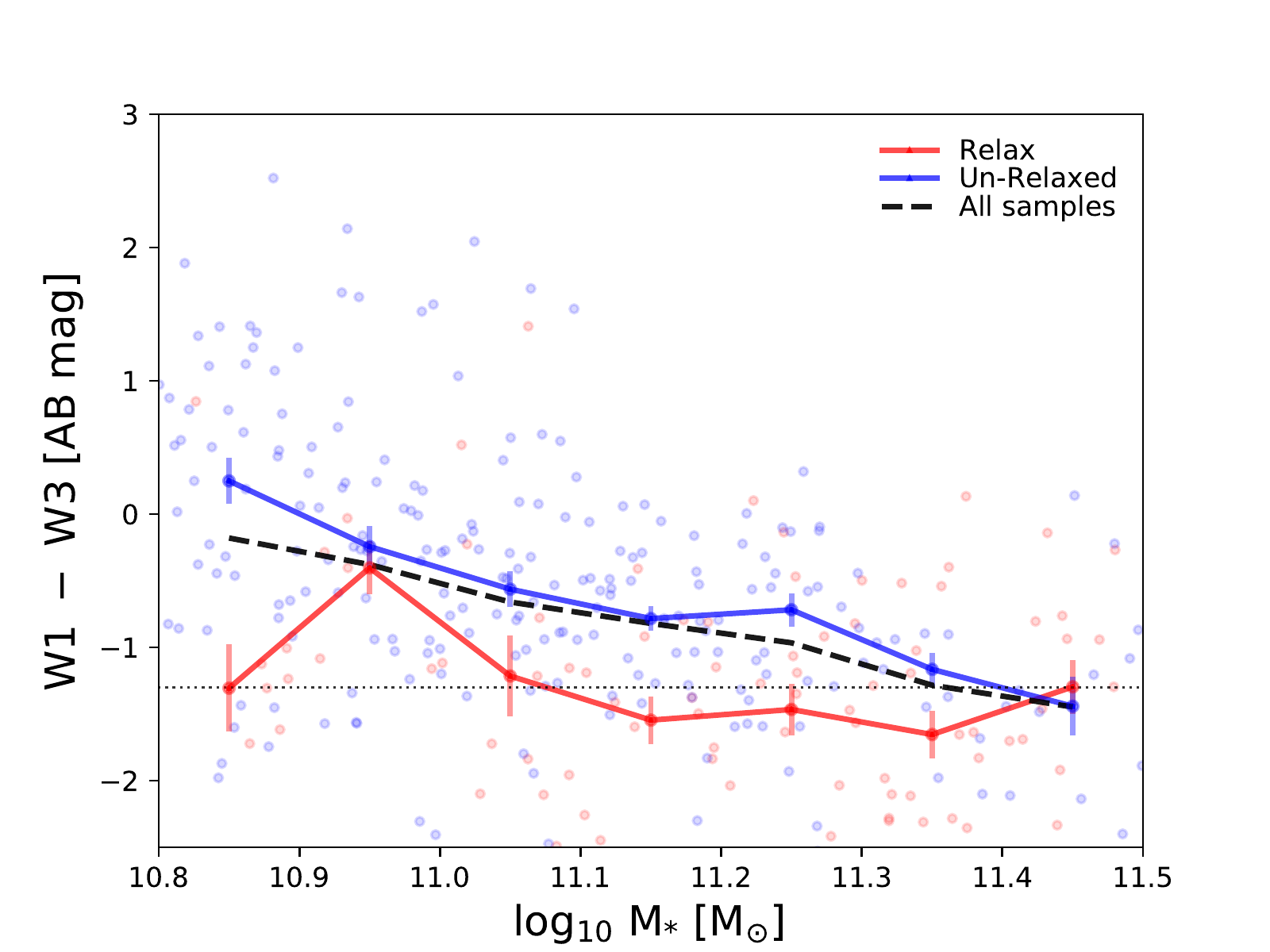}
		\caption{The metallicity(top panels), sSFR (middle panels) and w1-w3 color of {\sc{WISE}} mid-IR data as a function of stellar mass for relaxed/unrelaxed (right), low/high offset (middle) and high/low luminosity gap (left) with the medians and $\sigma/\sqrt{N}$ uncertainties. The dashed black line in each panel is the metallicity and sSFR -- stellar mass relation for the full sample. The region below the dotted line in the bottom panels shows where quiescent galaxies are expected to be found as described in \cite{Ko2013}.
		}
		\label{fig:nuv-metal-ms-am}
	\end{figure*}
	
	\subsection{Metallicity and sSFR}
	Figure \ref{fig:nuv-metal-ms-am} shows the metallicity and sSFR derived from SED-fitting as a function of stellar mass for all the sub-samples. The top panels of the figure show the relaxed systems have a lower ($\sim$ 0.05 dex) metallicity compare to the unrelaxed systems.
	There is no significant difference in the BGG metallicity of high and low offset systems at a given stellar mass. However, the luminosity gap panel appears very similar to the relaxed/unrelaxed systems panel. 
	
	In the middle row of the figure, the median sSFR of BGGs hosted by unrelaxed groups is higher than BGGs in relaxed groups. Although the SFRs estimated by magphys cannot reach zero even in fully quenched galaxies, the point we wish to highlight is the presence of a clear difference between the subsamples. In this case, the difference is driven by both the luminosity gap and the BGG offset. Comparing with the median trend of the total sample, the BGGs in relaxed and unrelaxed groups tend to be a little below and above the `All samples' trend, respectively, both in metallicity and sSFR at a given stellar mass. Note that the result doesn't change much when we exclude the galaxies without FIR from the sub-samples. Note that there is no noticeable difference in the sSFR between two samples at high BGG stellar masses. 
	
	We conduct further analysis using near IR data from the {\sc{WISE}} survey in the bottom row of Figure \ref{fig:nuv-metal-ms-am}. The w1-w3 color is expected to be sensitive to star formation activity over the last 2 Gyrs \citep{Ko2013}. Our results show higher w1-w3 values for the BGGs in unrelaxed systems, meaning they are less quiescent compared to the BGGs in relaxed groups.
	
	\subsection{Selection of AGN-host galaxies: BPT and WISE color-color diagrams}
\begin{figure*}
		\centering
		\includegraphics[width=0.33\linewidth]{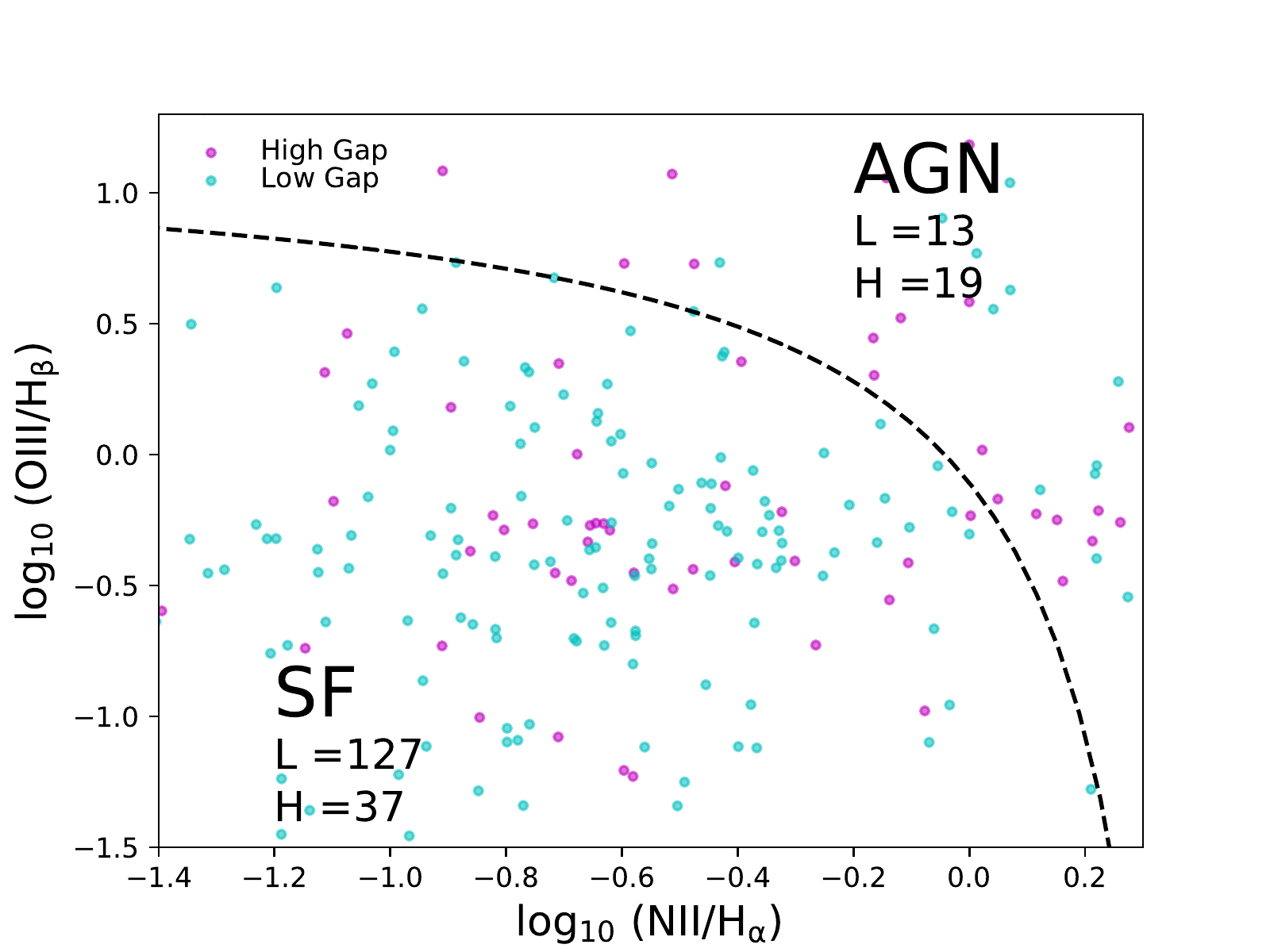}
		\includegraphics[width=0.33\linewidth]{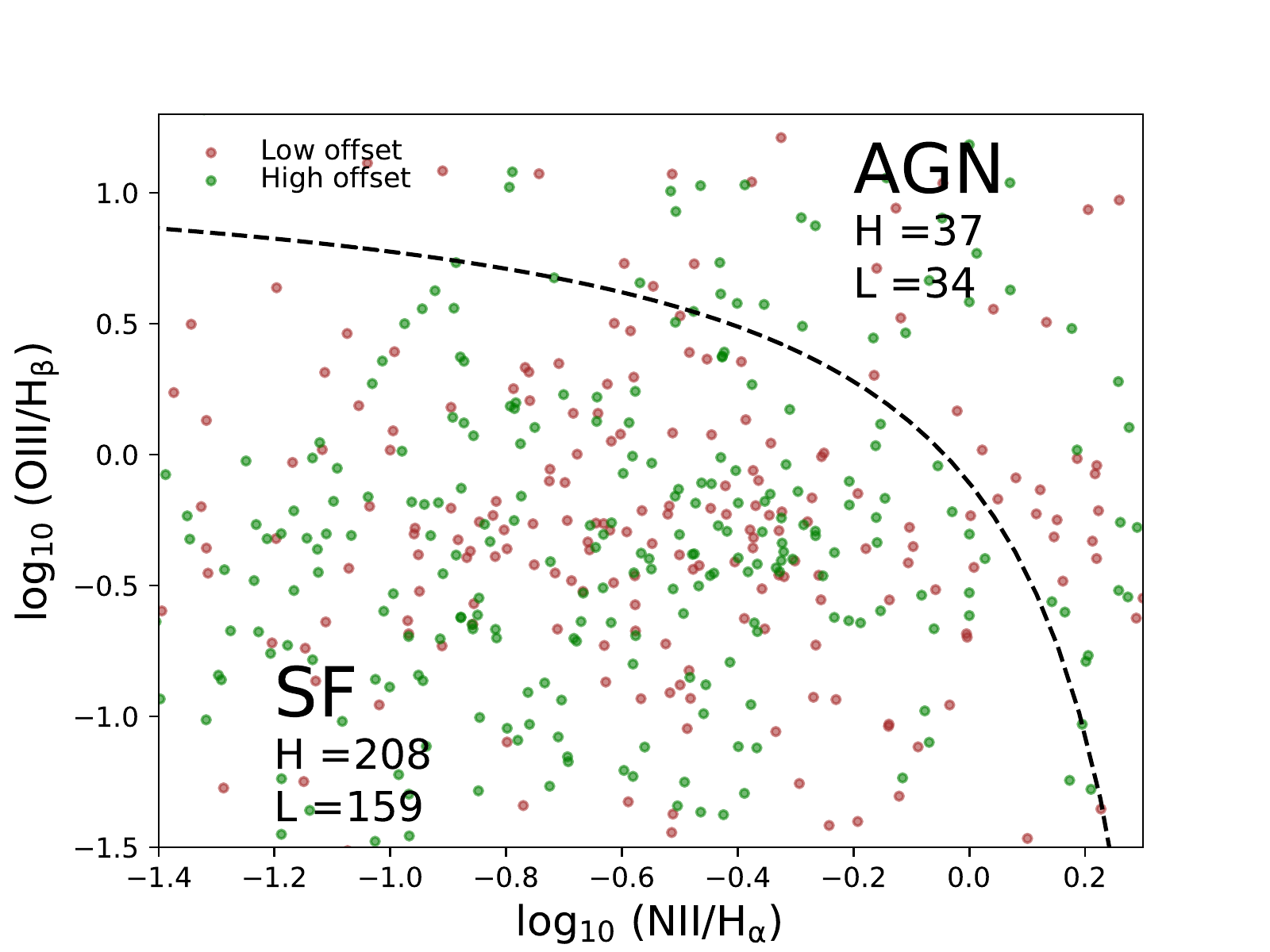}
		\includegraphics[width=0.33\linewidth]{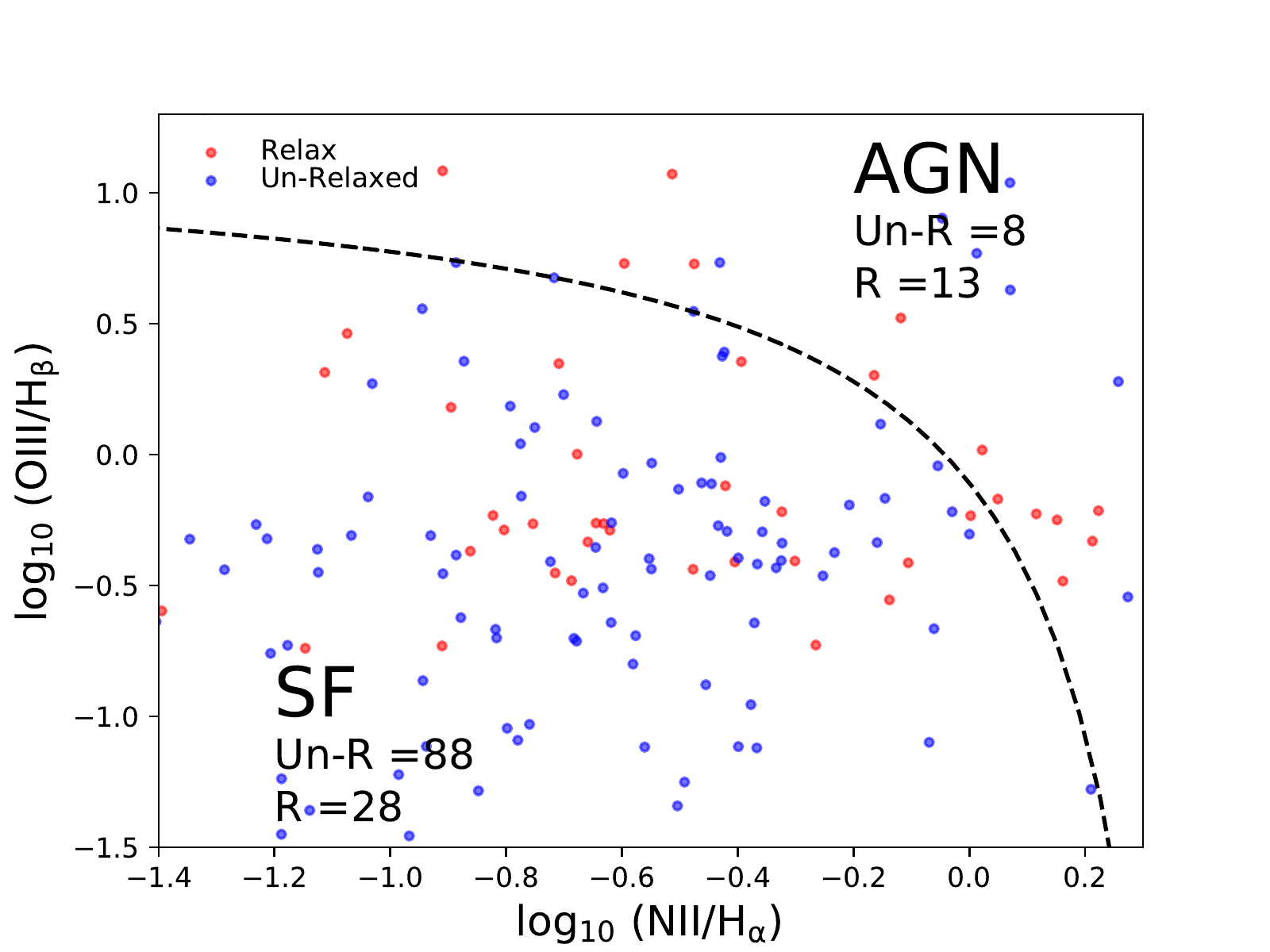}
		\includegraphics[width=0.33\linewidth]{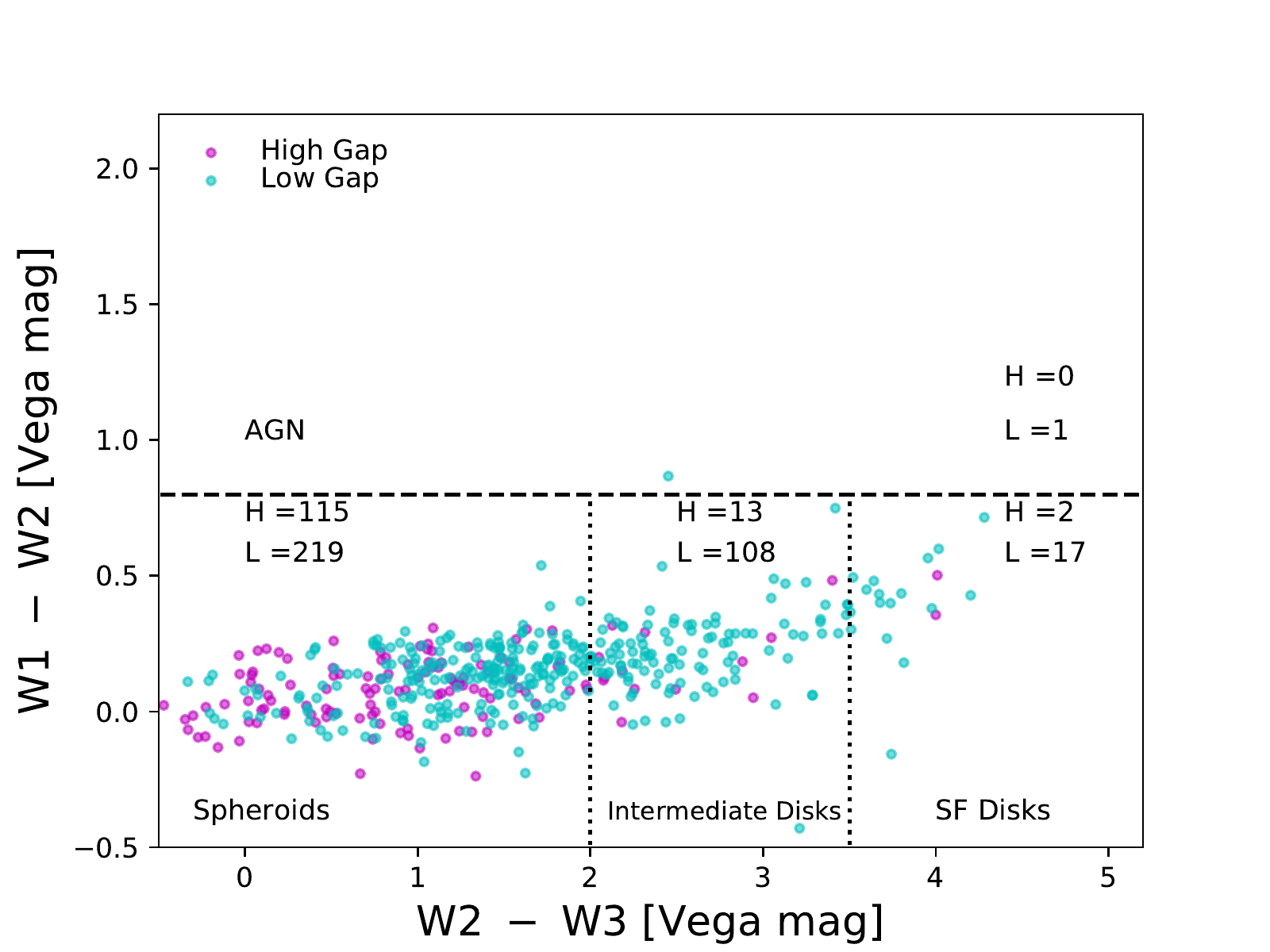}
		\includegraphics[width=0.33\linewidth]{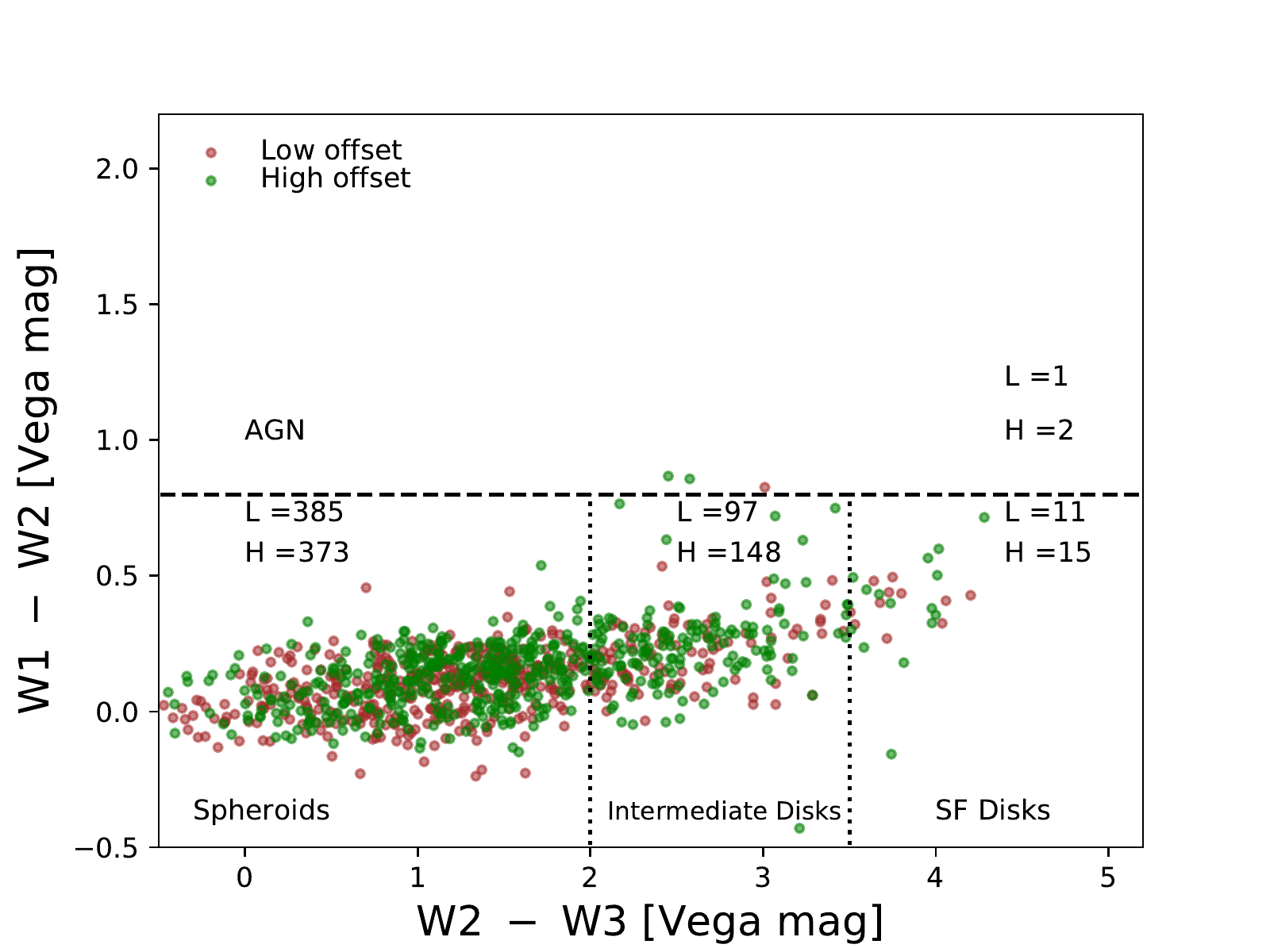}
		\includegraphics[width=0.33\linewidth]{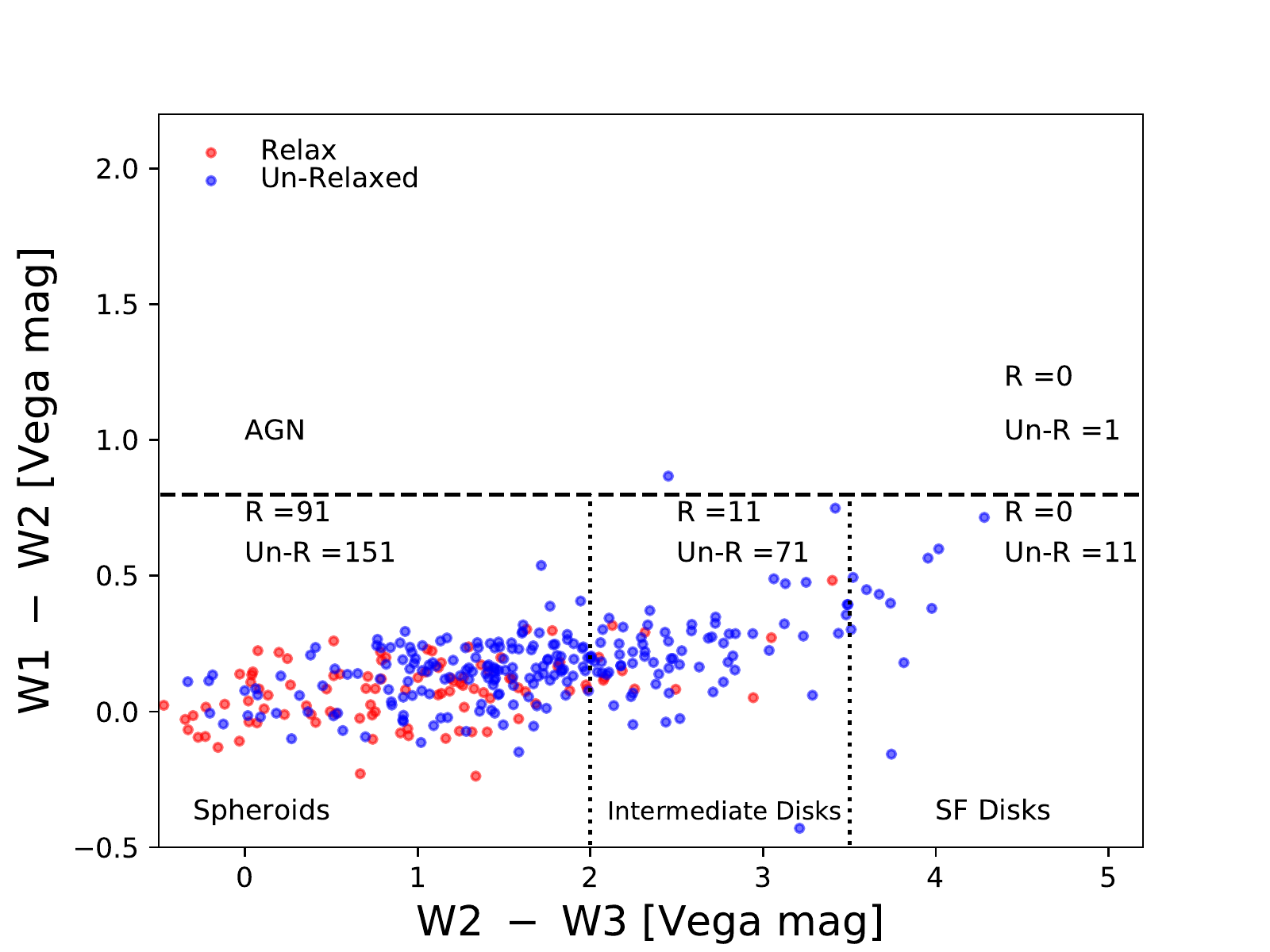}
		\caption{Top: Distribution of $OIII/H_{\beta}$ as function of $NII/H_{\alpha}$, BPT diagram, for the BGG in relaxed(R)/unrelaxed(Un), high(H)/low(L) gap and low(L)/high(H) offset sub-samples with categorizing the region and statistics of AGN and star forming galaxies from \citet{Kewley2001} show by dotted line. Bottom: The {\sc{wise}} color-color distribution of w1-w2 a function of w2-w3 (Vega mag) for the same top panel sub-samples. The AGN and star forming galaxies categorizing by the dashed line based on \citet{Stern2012}. The two dotted lines show the regions dominated by different morphological types including spheroids, intermediate and star forming disks. Each panel includes the number statistics of each sub-sample in that regions.}
		\label{fig:nuv-BPT}
	\end{figure*}

	The star formation rates and metallicities in this study were derived by {\sc{MAGPHYS}} SED fitting, which does not include modelling for the effects of AGN. Thus, here we attempt to exclude galaxies with AGN from our sample to check if they have influenced our results. In Figure \ref{fig:nuv-BPT} we show the BPT \citep{Baldwin1981} diagram (top panels) for all our sub-samples. In the bottom row, we use an alternative method to identify AGN in our sample, the w1-w2 vs. w2-w3 color diagram \citep{Jarrett2017,Cluver2014}. Combining the two methods to identify AGN, we find that less than 10\% of our samples are identified as AGN. In Figure \ref{fig:nuv-metal-ms-am}, the faint dashed lines show there is no significant impact on our star formation rate or metallicity results if we exclude those BGGs identified as AGN from our sample.

	Further, The morphology of the disk-like galaxies is also consistent with our elliptical probability in Figure \ref{fig:pellipgama} that illustrates the higher fraction of the BGGs in the unrelaxed groups tend to be disk-type galaxies in comparison to the BGG in the relaxed groups.
	
	\subsection{Velocity offset  indicator}
\begin{figure}
		\centering
		\includegraphics[width=0.9\linewidth]{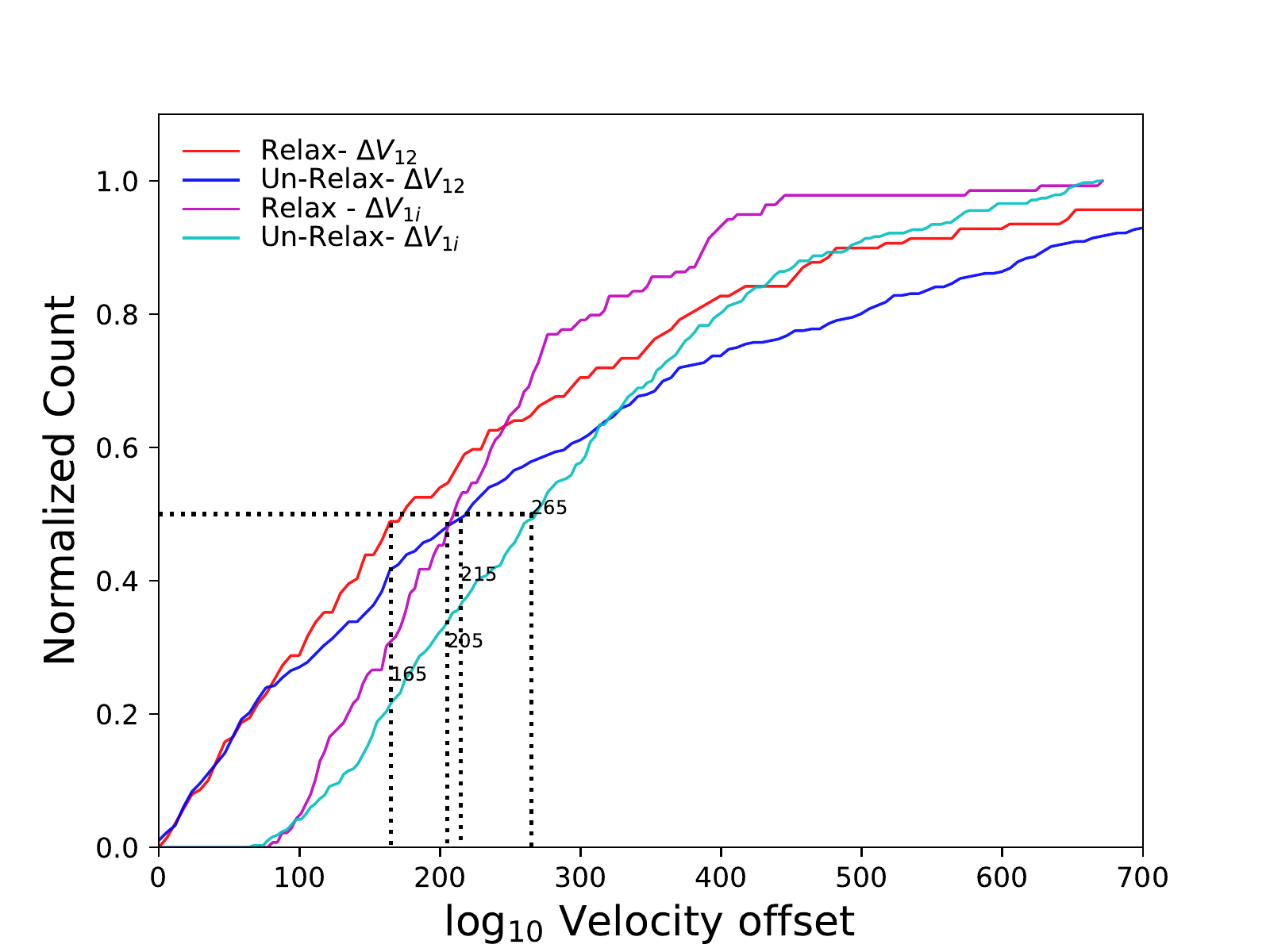}
		\caption{Cumulative distribution of the velocity offset between the 1st and 2nd most massive galaxy ($\Delta V_{12}$) and the 1st and $i$th most massive galaxy $\Delta V_{1i}$ galaxies within $0.5 R_{vir}$ for the relaxed and unrelaxed groups. The dotted lines and percentage show the median velocity offset for each sub-sample}
		\label{fig:nuv-hist-velgap-ru}
	\end{figure}

	Our approach to identifying relaxed and unrelaxed groups in this study does not directly require measurements of galaxy dynamics, meaning it can be more easily applied to large samples of groups. However, it might be expected that unrelaxed groups would show larger velocity offsets between BGGs and their groups. We investigate this velocity offset by comparing our relaxed and unrelaxed subsamples. We calculate the velocity offset in two ways; (i) between first and second brightest galaxies in each groups within 0.5 $R_{vir}$ ($\Delta V_{12}$), and (ii) between the first and the $i$th spectroscopic galaxies ($\Delta V_{1i}$) within 0.5 $R_{vir}$. With more spectroscopically confirmed members (i.e., method (ii)), the group means velocity is expected to be more accurately measured. Figure \ref{fig:nuv-hist-velgap-ru} shows the cumulative distribution of the velocity offset for the relaxed and unrelaxed groups of galaxies. The velocity offsets for the unrelaxed groups are always higher than in the relaxed groups of galaxies as was expected. The median values with SD error are $\Delta V_{12}$=165$\pm$17.5 km/s for relaxed groups, compared to 205$\pm$13 km/s for unrelaxed groups. The median of $\Delta V_{1i}$ is 205$\pm$9.2 km/s for the relaxed groups compared to 265$\pm$8.7 km/s for the unrelaxed groups. We also confirm this result further by selecting only groups with at least 4 members within the half virial radius and get very similar results to the $\Delta V_{1i}$ results.  We suggest that the velocity offset parameter for galaxy groups is likely an important dynamical age indicator for use in spectroscopic surveys. We will present further analysis of this indicator in future studies.
	
	\section{Discussion and conclusions}
	Using a sample of galaxy groups from the GAMA survey we study the stellar population properties and evolution of the brightest galaxies in the groups with different dynamical states namely relaxed and unrelaxed. We use two independent indicators to probe the dynamical state of the halo: the magnitude gap between the two brightest galaxies, $\Delta M_{12}$, and the offset present between the position of the BGG and group's luminosity weighted centre, $D_{offset}$.  We focus on BGGs in the groups which are more luminous than -21.5 in r-band magnitude in order to maximize the sample size for our redshift range, while ensuring completeness. In practice, this excludes more modest groups, containing Milky-Way like BGGs, from our sample.
	
	We find some clear differences between the BGGs in unrelaxed groups vs relaxed groups. There are higher numbers of blue BGGs (defined as having NUV-r$<$4.5) in the unrelaxed groups ($\sim$35\%) compared to the relaxed groups($\sim$14\%). In fact, they are bluer even at fixed Sersic index (our proxy for galaxy morphology), and at fixed dust mass, implying a difference in their recent star formation. We also find larger numbers of galaxies with non-elliptical morphology (disk-shaped with P(E) $<$ 0.5) in our unrelaxed sample ($\sim$38\%) with respect to the BGGs in our relaxed sample ($\sim$15\%), and they tend to have lower sersic indices.
	
	We compare our SED-fitting derived star formation rates and find unrelaxed groups show more star formation. This difference could partly be a result of the increased number of recent mergers expected from simulations in unrelaxed groups \citep[][see also \cite{Hwang2009} for similar results in observations]{Raouf2018}. If some of these are wet mergers, then it could provide additional fuel for the observed star formation. It is also clear in the optical r-band imaging of some representative galaxies, shown in Figure \ref{fig:Unrelax},  that the unrelaxed systems are much more crowded and irregular fields, which could provide greater opportunity for increased numbers of mergers. We also note that the higher star formation rates could also be a result of larger numbers of more disk-like BGGs in the unrelaxed groups. There are more low sersic-index galaxies (see right panels of Figure \ref{fig:gap_distribution}), and lower probabilities of finding elliptical morphology BGGs (left panel of Figure \ref{fig:pellipgama}) in the unrelaxed groups. However, given that we see bluer colours even at fixed sersic index (Figure \ref{fig:NUV_flux_fossil_ctrl_scater}), we conclude that at least some of the difference in colours and SFRs is probably as a result of increased numbers of mergers.
	
	We also compare the SED-fitting derived stellar metallicities of BGGs in relaxed and unrelaxed groups. We find that the metallicity of unrelaxed groups tends to be higher. This might be as a result of the mass of building blocks from which they formed. Groups that formed recently might be built from more massive and more metal-rich building blocks that have had longer to enrich themselves than the early formed groups. In \cite{Raouf2018}, using the {\sc radio-SAGE} model, we showed that BGGs in dynamically unrelaxed groups have suffered their last major galaxy merger typically $\sim 2$ Gyr more recently than BGGs in dynamically relaxed groups, and that this impacts on the black hole growth and activity \citep[e.g., see merger rates in fig. 2 of ][]{Raouf2018}. 
	
	Our general findings in this study might appear in contrast with the results of an earlier study by \cite{Trevisan2017}, who found no clear sign of the correlation between the luminosity gap and the ages, metallicities, [$\alpha$/Fe], and SFHs of BGGs using the sample of SDSS galaxy groups with elliptical BGGs. They also suggested that the BGG in high gap groups underwent dry mergers at an early stage as they find no trend between the $\Delta M_{12}$  and the BGG SFH. The two studies adopt two completely different approaches. The key difference between the two studies is the techniques employed for probing the star formation history. While we use an SED fitting technique, thus enabling us to incorporate the galaxy in its entirety, the study by \cite{Trevisan2017} is based on the SDSS fiber spectroscopic data which probes the central region of galaxies, due to the fiber diameter and positioning. Given the distance of the sample galaxies in \cite{Trevisan2017}, the diameter covered by their technique is limited to $\sim$ 4 kpc which is only a fraction of the size of these generally giant elliptical galaxies. In addition, the \cite{Trevisan2017} sample were selected from Galaxy Zoo \cite{Lintott2011} to be ellipticals only. This might have selected against some of the bluer BGGs in our sample, especially for objects with early-type morphology but blue colours. Our sample does not have a morphological restriction, and indeed we do see that unrelaxed BGGs appear bluer at a fixed value of sersic index. Furthermore, as the SFR depends on the morphological type and stellar mass and different redshift limit due to survey completeness, the two samples adopted by this study and that of  \cite{Trevisan2017} will present different characteristics. Considering the deeper imaging and higher redshift range of the GAMA sample with respect to the SDSS survey, we find more than 20\% of our sample have $\Delta M_{12} > 2.5$. The \cite{Trevisan2017} sample were restricted to $\Delta M_{12} \le 2.5$. This means we include even higher gap systems in our sample, which predominantly impacts on the results of our relaxed subsample. Last but not least, by using the SED fitting we consider their FIR and NIR emission which allows us to partly correct for dust extinction, and therefore our study should be less biased in determining SFRs compared to \cite{Trevisan2017}, who rely on spectroscopic observations limited to visible wavelengths. Given this large variety of differences in sample and approach, we believe it is not too surprising if differences are seen between our two studies in terms of star formation measurements.

	Finally, we find that for most of the differences found between the relaxed and unrelaxed groups, the magnitude gap, $\Delta M_{12}$, is the most important factor in driving the differences. The one exception is in the sSFR main-sequence where differences between the relaxed and unrelaxed groups are driven quite equally by both the luminosity gap and BGG offset. We check for the impact of AGN on our results by identifying AGN in our sample using a BPT diagram and WISE colour diagram (Figure \ref{fig:nuv-BPT}) and find that less than 10\% of our sample present as AGN. We confirm that our results do not change significantly when we exclude the identified AGN from our sample in Figure \ref{fig:nuv-metal-ms-am}. Finally, we find that groups identified as unrelaxed using our criteria show higher velocity offsets between their BGGs and host group.
	
	In our future studies, we will focus on how the dynamical state of the group impacts on the gas and stellar kinematics in group BGGs using IFU data.
	
	\acknowledgments
	We thank the referee for constructive comments and suggestions which helped to improve the paper.
	GAMA is a joint European-Australasian project based around a spectroscopic campaign using the Anglo-Australian Telescope. The GAMA input catalog is based on data taken from the Sloan Digital Sky Survey and the UKIRT Infrared Deep Sky Survey. Complementary imaging of the GAMA regions is being obtained by a number of independent survey programs including GALEX MIS, VST KiDS, VISTA VIKING, WISE, Herschel-ATLAS, GMRT and ASKAP providing UV to radio coverage. GAMA is funded by the STFC (UK), the ARC (Australia), the AAO, and the participating institutions. The GAMA website is http://www.gama-survey.org/.
	We would like to acknowledge financial support from ICRAR, AAO, ARC, STFC, RS, and ERS for GAMA Panchromatic Swarp Imager (PSI). MR benefited from discussions with Gary Mamon and Jae-Woo kim in this study.
	

\end{document}